\documentclass[aps,nofootinbib]{revtex4}
\usepackage[dvipsnames,usenames]{color}
\usepackage{amsmath}
\usepackage{graphicx}
\def\lsim{\mathrel{\raise3pt\hbox to 8pt{\raise -6pt\hbox{$\sim$}\hss{$<$}}}}
\def\rsim{\mathrel{\raise2pt\hbox to 9pt{\raise -7pt\hbox{$\sim$}\hss{$>$}}}}
\def\zrlow{\raisebox{-0.6ex}{\scriptsize \rm zr}}

\def\sstar{\raisebox{1.0ex}{\scriptsize *}}
\def\hip{\raisebox{1.0ex}{\scriptsize \it p}}

\def\nr2{\raisebox{-0.5ex}{\scriptsize \it n}}

\def\haf{\textstyle{1\over2}}

\def\vr{{\bf r}}

\def\vx{{\bf x}}

\def\vy{{\bf y}}
\def\vz{{\bf z}}
\def\vk{{\bf k}}
\def\vp{{\bf p}}
\def\vq{{\bf q}}
\def\vA{{\bf A}}
\def\vD{{\bf D}}
\def\vO{{\bf O}}
\def\vJ{{\bf J}}
\def\vL{{\bf L}}

\def\vsig{\mbox{\boldmath$\sigma$}}
\def\vmu{\mbox{\boldmath$\mu$}}
\def\valf{\mbox{\boldmath$\alpha$}}

\newskip\humongous \humongous=0pt plus 1000pt minus 1000pt
\def\caja{\mathsurround=0pt}

\newif\ifdtup
\def\panorama{\global\dtuptrue \openup1\jot \caja
        \everycr{\noalign{\ifdtup \global\dtupfalse
        \vskip-\lineskiplimit \vskip\normallineskiplimit
        \else \penalty\interdisplaylinepenalty \fi}}}
\def\eqalignno#1{\panorama \tabskip=\humongous
        \halign to\displaywidth{\hfil$\displaystyle{##}$
        \tabskip=0pt&$\displaystyle{{}##}$\hfil
        \tabskip=\humongous&\llap{$##$}\tabskip=0pt
        \crcr#1\crcr}}
\begin{document}
\hspace*{-.5in}
\rotatebox{90}{%
\fbox{\parbox[t]{1.03in}{LA-UR-13-23756}}
}
\vspace*{-1.05in}
\vspace*{0.5in}
\title{Nuclear Polarization Corrections to $\mu$-d\\
Atoms in Zero-Range Approximation}

\author{J.\ L.\ Friar\footnote{Electronic address: friarjim@aol.com}}
\affiliation{Theoretical Division,
Los Alamos National Laboratory,
Los Alamos, NM  87545}

\begin{abstract}
Nuclear polarization corrections  to the 2P-2S Lamb shift in $\mu$-d atoms are developed in order ($\alpha^5$), and are shown to agree with a recent calculation. The nuclear physics in the resulting corrections is then evaluated in zero-range approximation. The dominant part of the correction is very simple in form and differs from a recent potential model calculation by less than 1\%. It is also demonstrated how the third Zemach moment contribution largely cancels against part of the polarization correction, as it did in e-d atoms and does so exactly for point-like nucleons. This suggests that it may be possible to reduce the uncertainty in the theory (of which nuclear polarization is  the largest contributor) to less than 1\%.
\end{abstract}

\maketitle

\pagebreak
\section{Introduction}

The recent measurement of the proton root-mean-square charge radius at PSI \cite{PSI} has been an exciting  development in atomic and low-energy hadronic physics. Using the 2P-2S Lamb shift in $\mu$-p atoms the resulting $\langle r^2 \rangle^{1/2}_{\rm ch}$ = 0.84087(39) fm is about 7 standard deviations smaller than the value of the same quantity deduced from measurements of electron interactions with the proton. The latter can be obtained either from electron-proton scattering or the Lamb shift in e-p atoms; results from those experiments have been combined and summarized in CODATA-2010 \cite{CODATA}. Thus far no compelling explanation of the large discrepancy exists, although the possibility of new physics is a consideration. In an effort to crosscheck this result other experiments have been performed (the PSI $\mu$-d Lamb shift experiment is currently undergoing analysis \cite{franz}) and experiments in $\mu$-He atoms are planned. 

Nuclear polarization contributions  to the Lamb shift are among the most difficult corrections to calculate accurately and can be quite large. These are dynamic contributions to energy levels, and can be viewed naively as the Coulomb attraction of the lepton pulling the protons in a nucleus away from the nuclear center-of-mass (CM). The distorted charge distribution then tries to follow the lepton in its orbit around the nucleus, much like the tides in the Earth-Moon system follow the Moon, and this lowers the overall energy. This cartoon description clearly indicates that only excited states of the nucleus contribute (they reflect the distortion), while emphasizing the role of the dominant electric dipole excitations in a nucleus. Indeed, the giant dipole resonance can be naively viewed as a simple oscillation of the nuclear protons against the neutrons \cite{GDR}.

A recent calculation \cite{KP} of the polarization corrections for the $\mu$-d experiment currently undergoing analysis found a relatively large contribution dominated by (virtual) dipole excitations. Given the complexity of the nuclear force models required to calculate these corrections, a reasonable question is the size of the uncertainty given the underlying physics, even if the numerical precision is exact. We will try in this work to provide insight into this question, not by performing another nearly identical numerical calculation, but by performing an approximate treatment that emphasizes some of the unique properties of the deuteron. The weak binding of the deuteron and the relatively short range of the nuclear force motivated Bethe and Peierls \cite{zero} to develop the zero-range approximation, which circumvented the nearly complete lack of information about the nuclear force at the time of that work. We will follow that approach, which has numerous advantages that we summarize as follows.

Our motivation for a new polarization calculation is that it provides:
(1) an alternative treatment of basic formulae that we express as energy-weighted photonuclear sum rules;
(2) an alternative and greatly simplified treatment of the nuclear physics;
(3) estimates of neglected terms;
(4) an independent framework for treating nuclear polarization in $\mu$-He atoms \cite{myhe4};
(5) a common framework with previous treatments of the e-d atom \cite{higher}.

Why are we interested in performing an {\sl approximate} treatment of polarization corrections, when a more accurate calculation \cite{KP} already exists? A zero-range approximation \cite{zero} calculation will:
(6) produce simple and quite accurate formulae for all contributions based only on one- and two-nucleon observables;
(7) allow uncertainty estimates based on uncertainties in these observables ($\sim 0.2$\%);
(8) test sensitivity of the results to the interior part of the deuteron wave function, which is determined by details of the nuclear force;
(9) allow independent estimates of total error ($\lsim 1$\%);
(10) allow us to interpret many of our results in simple terms. The leading-order polarization correction in zero-range approximation will be shown below to differ from the corresponding complete calculation of Ref.~\cite{KP} by only 0.9\%. This small error reflects the influence of the (quite complicated) nuclear force on the {\sl interior} part of the deuteron wave function, which part is obviously  not very important in the final result. This insensitivity to details is the {\it raison d'etre} of the zero-range approximation.

Item (9) in the list above is particularly {\it apropos} if the uncertainties in the PSI $\mu$-d experiment  and their $\mu$-p experiment (viz.,  $\epsilon_p \sim$ 0.004~meV) are comparable. The estimated uncertainty in the polarization corrections of Ref.~\cite{KP} was 0.016 meV,  or $4\, \epsilon_p$. Any insight into theoretical uncertainties is likely to be valuable. We note that the deuteron is in all likelihood the only nuclear case where such uncertainties can be lower than 1\% \cite{d-pol}, and this is entirely due to its weak binding.

Our organization of this manuscript is unusual, which reflects in part a desire to present a rather comprehensive and self-contained treatment of the zero-range approximation \cite{zero}, the latter being both very useful and underused. This treatment occupies  Appendix B.  In order to facilitate a pedagogical approach we have banished the most complicated remaining mathematical details to other appendices, and treat only the most significant aspects in the main body of the manuscript, including numerical estimates.

We begin in Section~(2) by deriving the basic formulae for the polarization corrections in muonic atoms, using an approach previously developed for electronic atoms \cite{higher}. We argue that a simple non-relativistic approach to the atomic physics yields the dominant contributions. This is explicitly demonstrated by performing the rather complicated relativistic (i.e., Dirac muon) calculations in Appendix A, where additional (but quite small) contributions are calculated. We then present in Section~(3) numerical results for each of our individual terms in zero-range approximation and compare them to the accurate results of Ref.~\cite{KP}. We conclude in Section~(4). In Appendix B we present an introduction to zero-range theory and perform all of the calculations of our zero-range terms. The calculational part of this appendix is largely mathematical, and may be of little interest to some readers, but it is self-contained and complete. Appendix C contains some additional mathematical details for the interested reader. Note that we work in natural units with $\hbar = c =1$, and these quantities must be inserted in our formulae in order to obtained usable formulae in other units.

\section{Polarization Corrections}

We require the energy shift in the $n$S state of a muon of mass $m$ interacting with a nucleus of mass $m_t$ via two-photon exchange in their CM frame. This is conveniently expressed in leading order in $\alpha$ (viz., $\alpha^5$) in terms of their at-rest forward scattering amplitude, as discussed in Ref.~\cite{myhe4}. Each exchanged photon then has four-momenta $q^{\mu}$ that are equal in magnitude and oppositely directed as indicated in Fig.~(1a). Because the muon mass is much greater than the other energies in the problem, the muon moves slowly in the intermediate state and this generates very small electromagnetic currents (which is the opposite of the e-d problem). Consequently the dominant terms (by far) are the interactions of the muon charge with the deuteron charge. This dominance is conveniently highlighted by using Coulomb gauge in the calculation, which leads to ordinary static Coulomb interactions between the charges. Consequently only the transverse parts of the current and seagull contributions (the latter shown in Fig.~(1c)) are required. The seagull amplitude is required for gauge invariance, which is necessary in order to obtain finite results and useful for simplifying those results (see Appendix B of Ref.~\cite{higher}). Because we are only interested here in the {\sl inelastic} nuclear processes (viz., virtual excitations that produce the polarization corrections), gauge invariance requires us to use only the {\sl inelastic} part of the seagull amplitude. Note that the elastic part of the seagull amplitude (which we will not treat) generates recoil corrections and is nuclear-structure dependent. We will also refer below to static contributions that require only the deuteron ground state for their calculation as ``elastic'' contributions, in contradistinction to the inelastic ones we develop herein.

\begin{figure}[htb]
\includegraphics[width=5.3in]{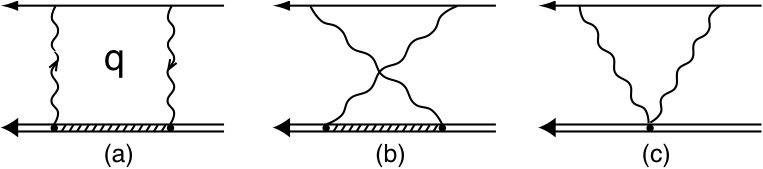}
\caption{Nuclear Compton amplitude with direct (a), crossed (b), and seagull (c)
contributions illustrated. Single lines represent a muon, wiggly lines a photon propagator (with four-momentum $q^{\mu}$), unshaded double lines a nuclear ground state,  while 
shaded double lines depict a nuclear Green's function containing a sum over
{\sl excited} nuclear states. The seagull vertex in (c) maintains gauge invariance by incorporating the effect of ``frozen'' nuclear degrees of freedom, such as nucleon-antinucleon pairs or pions \cite{review}.}
\end{figure}

We argued in the Introduction that non-relativistic physics is dominant for the $\mu$-d atom. This was our approach in Ref.~\cite{myhe4} for the $\mu-$He atom. We will verify that dominance in Appendix A by calculating the complete set of corrections of appropriate size. In addition to the dominant set of non-relativistic terms, one additional small but non-negligible term of relativistic origin is found, which we also derive below via a simple modification of our non-relativistic formalism  \cite{myhe4}. Only one new term of marginal importance is found in Appendix A, plus neglected terms that we introduce and estimate in Section~(3). The interaction of the muon electromagnetic current with the nuclear current produces that marginal term, and we therefore ignore until later the effect of the current-current interactions. The rather small seagull terms are primarily required to enforce gauge invariance of the nuclear currents, and can therefore also be ignored here (but will be incorporated in Appendix A and estimated in Appendix B). 

In Coulomb gauge for non-relativistic muons the contribution of Fig.~(1b) vanishes for the interactions between charges (virtual muon-pair intermediate states required by relativity are the primary contribution). Thus we only need to calculate the contribution of Fig.~(1a), which precisely equates to the muon and deuteron charges interacting via static Coulomb potentials.

The (attractive) energy shift for the $n$\underline{th} S-state of the atom due to nuclear polarization is given to leading order in the fine-structure constant $\alpha$ by
$$
\Delta E_{\rm pol}^{\rm NR} = - 8 \alpha^2 |\phi_n (0)|^2 \sum_{N \neq 0}  \int  \frac{d^3 q}{4 \pi} \frac{ \; \langle 0 | \rho_{\rm ch} (-\vq) | N \rangle \langle N | \rho_{\rm ch} (\vq) |0 \rangle }{q^2\;(\omega_N + \frac{q^2}{2\, m_r})\, q^2}\, . \eqno(1)
$$
This is nothing more than ordinary second-order perturbation theory in non-relativistic quantum mechanics for an energy shift in configuration-space that has been rewritten in momentum space (and derived as Eqn.~(7) in Ref.~\cite{myhe4}). 

The (virtual) nuclear excitations driven by the muon are localized inside the nucleus at the center of the atom, which accounts for the factor of $|\phi_n (0)|^2 = (Z \alpha\, m_r/ n)^3 /\pi\,$, the square of the muon wave function at the nucleus for the $n$\underline{th} S-state. Note that  the deuteron has charge $Z=1$  and that $m_r$ is the usual $\mu$-d reduced mass formed from $m$ and $m_t$. In the energy denominator $\omega_N = E_N - E_0$ is the energy difference between the $N$\underline{th} intermediate (excited) state ($|N \rangle$) of the deuteron and its ground state ($|0 \rangle$),  while $q^2/2\, m_r$ is the kinetic energy difference in the atom of the intermediate state  and the ground state (which has none to leading order in $\alpha$). Two factors of $- 4 \pi \alpha\, \rho_{\rm ch}(\vq)/q^2$ arise from the Fourier transform of the static Coulomb interaction between muon and deuteron, while $\rho_{\rm ch}(\vq)$ is the Fourier transform of the deuteron's charge operator in configuration space: $\rho_{\rm ch} (\vq) = \int d^3x\, \exp{(i \vq \cdot \vx)}\, \rho_{\rm ch} (\vx)$. The usual  phase space factor of $1/(2 \pi)^3$ accompanies $d^3 q$, and with that inclusion all numerical factors in Eqn.~(1) are accounted for.

Moving the Fourier exponentials from the two factors of $\rho_{\rm ch} (\vq)$ in Eqn.~(1) directly into the $\vq$-integral produces a much more tractable form
$$
\Delta E_{\rm pol}^{\rm NR} = - 8 \alpha^2 |\phi_n (0)|^2 \sum_{N \neq 0}  \int  d^3 x \int d^3 y\, \langle 0 | \rho_{\rm ch} (\vy) |N \rangle \langle N | \rho_{\rm ch} (\vx) |0 \rangle \; I_{\rm NR} (z) \, , \eqno(2a)
$$
where $\vz \equiv \vx - \vy$. All of the coupling between the atomic and nuclear physics is now contained in the structure function
$$
I_{\rm NR} (z) \equiv \frac{1}{4 \pi} \int \frac{d^3 q}{q^4} \frac{e^{i \vq \cdot \vz}}{\omega_N + \frac{q^2}{2 m_r}} = \frac{\lambda^2}{\omega_N z} \int_0^{\infty} \frac{d q}{q^3} \frac{\sin (q z)}{\lambda^2 + q^2}  \, , \eqno(2b)
$$
where we have defined $\lambda = \sqrt{2\, m_r\, \omega_N}$.  Changing integration variables to $q = \lambda\, t$ and defining $\beta = \lambda\, z$ then produces a simple result
$$
I_{\rm NR} (z)= \frac{1}{\lambda \,\omega_N\, \beta} \int_0^{\infty} \frac{d t}{t^3}\; \frac{\sin (\beta t)}{1 + t^2} = \frac{1}{\lambda \,\omega_N\, \beta}\, J_{\rm NR} (\beta)\, . \eqno(3)
$$
The dimensionless integral $J_{\rm NR} (\beta)$ in Eqn.~(3) diverges at small $t$. However, the small-$t$ limit of $\sin{(\beta t})$ contains a factor of $\beta$ that cancels an identical term in the prefactor of the integral. This term in the expansion is then independent of nuclear coordinates and thus incapable of exciting the nucleus. It therefore doesn't contribute to nuclear polarization and we  ignore it. The second term in the expansion of $\sin{(\beta t})$ is proportional to $\beta^3$ and is finite. This is the dominant term. The next term in the expansion would be proportional to $\beta^5$, but is linearly divergent, implying the existence of a $\beta^4$ term. Thus we have  $J_{\rm NR} (\beta) \sim a \beta^3 + b \beta^4 + c \beta^5 + \cdots$ . Equation~(3) will be the template for calculating most of the corrections that we require.

The simplest way to calculate $J_{\rm NR} (\beta)$ is to differentiate it twice and use identity {\sl 3.725.1} of Ref.~\cite{GR}
$$
J^{\prime \prime}_{\rm NR} (\beta) = - \int_0^{\infty} \frac{d t}{t}\; \frac{\sin (\beta t)}{1 + t^2} =  \frac{\pi}{2}(e^{-\beta} -1) \, . \eqno(4a)
$$
Straightforward integration then produces
$$
J_{\rm NR} (\beta) =   \frac{\pi}{2}( e^{-\beta} -1 +\beta - \beta^2 /2) \cong \frac{\pi}{2} \left( -\frac{\beta^3}{6} +\frac{\beta^4}{24} - \frac{\beta^5}{120} + \cdots  \right)\, ,  \eqno(4b)
$$
which agrees with the power series deduced above and generates
$$
I_{\rm NR} (z) = \frac{\pi}{2\lambda \,\omega_N\, \beta}(e^{-\beta} -1 +\beta - \beta^2 /2) \cong  \frac{\pi \lambda z^2}{6\,\omega_N}\left(- \frac{1}{2} +\frac{\beta }{8} - \frac{\beta^2}{40} + \cdots  \right)\, .  \eqno(5)
$$
After some manipulation we find our primary result
$$\eqalignno{
\Delta E_{\rm pol}^{\rm NR} =& - \frac{4 \pi}{3} \alpha^2 |\phi_n (0)|^2 \sum_{N \neq 0}  \int  d^3 x \int d^3 y\, \langle 0 | \rho_{\rm ch} (\vy) |N \rangle \langle N | \rho_{\rm ch} (\vx) |0 \rangle  \cr
&\left[ -\sqrt{\frac{2 m_r}{\omega_N}} \frac{z^2}{2} + m_r \frac{z^3}{4} - m_r^2 \sqrt{\frac{\omega_N}{2 m_r}} \frac{z^4}{10}  + \cdots \right]\, . &(6)
}
$$
These three terms with their proper dependence on the atomic reduced mass generate the bulk of the polarization corrections \cite{KP}. We will use the  scales in the problem to show below that each succeeding term in this series is roughly 1/4 (or less) of the preceding one. Since higher-order terms in the expansion become more and more sensitive to higher energies and the effect of relativity, not every term in Eqn.~(6) indicated by dots should be reproduced in the exact expression that we develop in Appendix A.

One additional small term of relativistic origin is easily derived using a simple modification of the above approach. We sketch the derivation performed  above Eqn.~(14$^{\prime}$) in Ref.~\cite{myhe4}, where more details can be found. In Eqn.~(1) we replace the  muon energy difference ($\vq^{\,2} /2 m_r$) in the energy denominator by the Dirac muon energy difference $(\valf \cdot \vq + \beta m_r) -m_r$ for a muon with reduced mass, $m_r$. This is the Breit approximation that generates a large tractable class of polarization corrections, and is discussed in detail in Ref.~\cite{breit} and in the {\sl Breit Approximation} subsection  at the end of our Appendix A. While the reduced-mass prescription isn't quite correct, such corrections aren't significant in the very small terms, and in this we follow Ref.~\cite{KP}. Rationalizing the energy denominator and using $\langle \valf \rangle =0$ and $\langle \beta \rangle=1$ simply redefines $\lambda$ to be $\lambda^{\prime} = \sqrt{2 m_r \omega_N (1 - \omega_N / 2 m_r})$ in Eqns.~(3) and (5). This is also identical to  the result from the Breit-approximation term in Eqn.~(A23). One finds that the leading term in Eqn.~(6) (i.e., $\sim z^2$) is multiplied by an approximate factor of $(1 - \omega_N/4 m_r)$. The second term in this factor is the leading-order correction due to relativity in the $\mu$-d atom.

In order to develop  tractable final expressions we manipulate the $z^n$ factors in Eqn.~(6) term-by-term, before collecting terms and writing the results. Squaring $\vz$ gives $z^2 = x^2 +y^2 -2 \vx \cdot \vy$, and we note that the $x^2$ and $y^2$ terms when inserted in Eqn.~(2a) don't contribute because one integral always becomes the total charge operator, which can't excite the deuteron. The remaining term generates two dipole operators since $\vD \equiv \int d^3 x\, \vx\, \rho_{\rm ch} (\vx)$.

The third term can be expanded in a similar fashion, and the results written as irreducible tensors in $\vx$ and $\vy$. This produces $z^4 \rightarrow 4 Q^{\alpha \beta}_x Q^{\alpha \beta}_y + \frac{10}{3} x^2 y^2 - 4 \vx \cdot \vy\, (y^2 + x^2)$, where we have dropped the $x^4$ and $y^4$ terms for the same reason we dropped the $x^2$ and $y^2$ terms above. We see that the $z^4$ term produces a quadrupole excitation term ($ Q^{\alpha \beta}$ is the quadrupole tensor), a monopole excitation term, and a retarded electric dipole term, respectively.

The remaining term ($\sim z^3$) is somewhat controversial, and is difficult to calculate {\it per se}. Simple expansions of $|\vx-\vy\,|^n$ for odd values of $n$  are not possible. These are the terms that lead to Zemach moments \cite{zemach,annals}. The relevant portion of Eqn.~(6) is
$$
 \int  d^3 x \int d^3 y\,\left [\sum_{N \neq 0}   \langle 0 | \rho_{\rm ch} (\vy) |N \rangle \langle N | \rho_{\rm ch} (\vx) |0 \rangle \right] \; |\vx - \vy|^3 \, . \eqno(7)
$$
Two significant features of this quantity are: (1) if one ignores the summation and replaces the states $|N \rangle$ by $|0 \rangle$, the usual third Zemach moment results; (2) since there are no energy factors inside the summation one can immediately use closure ($\sum_{N \neq 0} |N \rangle \langle N |  = 1 -  |0 \rangle \langle 0 |$) to rewrite Eqn.~(7) as
$$
 \int  d^3 x \int d^3 y\,\left (\, \langle 0 | \rho_{\rm ch} (\vy)  \rho_{\rm ch} (\vx) |0 \rangle  - \langle 0 | \rho_{\rm ch} (\vy) |0 \rangle \langle 0 | \rho_{\rm ch} (\vx) |0 \rangle \, \right) \; |\vx - \vy|^3 \, . \eqno(8)
$$
The subtracted term when integrated is the usual third Zemach moment of the deuteron: $\langle r^3 \rangle_{(2)}^{dd} \equiv  \int  d^3 x \int d^3 y\,\langle 0 | \rho_{\rm ch} (\vy) |0 \rangle \langle 0 | \rho_{\rm ch} (\vx) |0 \rangle\, |\vx - \vy|^3$. Using closure we can thus replace our {\sl inelastic} Zemach moment by the difference between a simple correlation function and an {\sl elastic} Zemach moment.

This is more relevant than just an independent technique for evaluating the inelastic Zemach moment. An elastic contribution to the Lamb shift exists in the form \cite{annals}
$$
\Delta E_{\rm el}^{\rm NR} = -\frac{4 \pi}{3} (Z \alpha)^2\, | \phi_n (0)|^2 
\left( \frac{m_r}{4} \langle r^3 \rangle_{(2)}^{dd} \right) \, , \eqno (9)
$$
which is equal and opposite to the second term  in Eqn.~(8) (when inserted in Eqn.~(6)) and exactly cancels it. This cancellation was originally demonstrated in Ref.~\cite{size} for e-d atoms, but holds equally well for $\mu$-d atoms \cite{KP}. The sum of the elastic and inelastic Zemach terms is therefore a deuteron charge correlation function that is  considerably simpler and significantly smaller than either Zemach term. We will use a compact notation for that function:
$$
\langle 0|\, |\vx-\vy|^3 | 0 \rangle_{\rm ch} \equiv \int  d^3 x \int d^3 y\, \langle 0\, | \rho_{\rm ch} (\vy)  \rho_{\rm ch} (\vx) | \, 0 \rangle \; |\vx - \vy|^3 \, . \eqno(10)
$$
This quantity is especially simple if the protons and neutrons are {\sl point-like}. Then all of the charge in the deuteron resides at a single point (on the deuteron's single proton) and $\vx - \vy$ vanishes, as does the correlation function. {\it In the point-nucleon limit the sum of the elastic and inelastic Zemach terms vanishes for the deuteron.} We note that it does not vanish for the He case (which has two protons) nor if the nucleons have finite size, a case that was not treated in Ref.~\cite{KP} but will be treated in detail in Appendix B. Because the sum of the inelastic and elastic terms is simpler to calculate and much smaller than either term, we follow Ref.~\cite{KP} and advocate using the sum.

The sum of the six largest nuclear polarization contributions (five of which are non-relativistic in origin) {\sl plus the elastic Zemach contribution} is therefore given by
$$\eqalignno{
\Delta E_{\rm pol} &= - \frac{4 \pi}{3} \alpha^2 |\phi_n (0)|^2 \left [ \sum_{N \neq 0} \sqrt{\frac{2 m_r}{\omega_N}} |\langle N | \vD |0 \rangle |^2 - \frac{1}{2} \sum_{N \neq 0} \sqrt{\frac{\omega_N}{2 m_r}} |\langle N | \vD |0 \rangle |^2\right . \cr
 &+ \frac{m_r}{4}\langle 0  | \,|\vx-\vy|^3 | 0 \rangle_{\rm ch} + \frac{4 m_r^2}{5}  \sum_{N \neq 0}\sqrt{\frac{\omega_N}{2 m_r}}|\langle N | \vD |0 \rangle\sstar \! \cdot  \langle N | \vO |0 \rangle   \cr
 &- \left . \frac{m_r^2}{3} \sum_{N \neq 0}\sqrt{\frac{\omega_N}{2 m_r}} |\langle N | \hat{r}^2 |0 \rangle |^2  -\frac{2 m_r^2}{5} \sum_{N \neq 0}\sqrt{\frac{\omega_N}{2 m_r}}  |\langle N | Q^{\alpha \beta} |0 \rangle |^2 \right] , & (11)
 }
$$
where we have introduced the monopole operator $\hat{r}^2$ (defined in Eqn.~(B5c)), the quadrupole operator $Q^{\alpha \beta}$ (defined in Eqn.~(B5d)), and the retarded dipole operator $\vO$ defined in Eqn.~(B5e). Because these multipoles arise from the charge operator (rather than the current operator) we denote them by C0, C2, and C1, respectively.
Equation~(11) is in complete agreement with Ref.~\cite{KP} except for the $\langle 0 |\, |\vx-\vy|^3 | 0 \rangle_{\rm ch}$ term, which vanishes with the point-nucleon assumption of that work. One additional small term resulting from magnetic spin-flip excitation  and derived in Appendix A will be included in the next section. Our result for this term also agrees with Ref.~\cite{KP}. We note that the first and second terms above were first derived in Ref.~\cite{myhe4}, while the final four terms were first derived in Ref.~\cite{KP}. Our results here completely agree with the latter work in the point-nucleon limit. Six additional smaller terms developed in Appendix A will be estimated in the subsection {\sl  Gauge Sum Rules} of Appendix B  and shown to be quite small.

We note in Eqn.~(5) that the sizes of the various terms in our expansion are determined by the dimensionless parameter $\beta = \sqrt{2 m_r c^2\, \omega_N}\, z/ \hbar c$, where we have reintroduced $\hbar$ and $c$ in the form $\hbar c$ = 197.327 MeV fm. We infer from the sum rules derived in Appendix B that the deuteron binding energy of $E_B$ = 2.225 MeV sets the energy scale for $\omega_N$ for the low-lying transitions that we require.  Then using $m_r c^2$ = 100.02 MeV and $z \sim R_d \sim$ 2 fm (the deuteron radius) produces the estimate $\beta \sim$ 1/4. Note that the parameter $m_r c^2\, R_d/\hbar c \sim 1$ does not help convergence, and the useful expansion parameter is therefore $\omega_N/m_r c^2 \sim 1/20$. This guarantees fairly rapid convergence of the series, which is further helped by the factorial-type convergence of the exponential. 

Nuclear current matrix elements are characterized by expansions in $Q/M_N$, where $Q$ is a typical momentum in the deuteron and equals $\kappa \sim$ 46 MeV in zero-range approximation (i.e., the virtual momentum in the tail of the deuteron wave function, which is discussed in detail in the next section). Current terms are thus reduced by $(\kappa/M_N)^2 \sim 1/400$ compared to typical charge terms. This makes almost all such terms negligible. The same argument applies to relativistic corrections in the deuteron, which are discussed at the very end of Appendix B.

\section{Zero-Range Approximation}

The zero-range approximation was developed in 1935 by Bethe and Peierls \cite{zero} in order to circumvent the almost complete lack of knowledge of the nuclear force at that time. It was known that the nuclear force had a range $R_V \sim 1$  fm, and that the deuteron was rather weakly bound. Weak binding means that the tail of the deuteron wave function extends well beyond that force. Outside the force the s-wave deuteron wave function is given by ($A_S\exp{(-\kappa\, r)}/\sqrt{4 \pi}\, r$). The amplitude  is determined by $A_S = 0.8845(8)\, {\rm fm}^{-1/2}$~\cite{deut}, the experimental deuteron s-wave asymptotic normalization constant, while  its extent is determined by the parameter $\kappa = \sqrt{2 \mu E_B}$ = 45.7022 MeV (or 0.23161 ${\rm fm}^{-1}$ after dividing by $\hbar c$).  This expression is determined by twice the n-p reduced mass, $2 \mu =  938.918$ MeV, and the deuteron binding energy, $E_B = 2.224575(9)$ MeV \cite{deut}. The $\kappa$-parameter corresponds to a deuteron length scale of approximately 4.3 fm, which is well outside the nuclear force. The dimensionless parameter $\kappa R_V \sim 1/4$ is therefore  reasonably small and typically occurs as  squared, cubed, and higher powers \cite{elpol} compared to a leading-order term $\sim 1$. Another parameter that can occur is $\kappa/M_N \sim 1/20$, where $M_N$ is the average nucleon mass that will be taken equal to $2 \mu$. These are all comfortably small parameters.

The essence of the zero-range approximation is to assume that the asymptotic form of the s-wave deuteron wave function holds everywhere. Matrix elements generated by integration then typically scale like $1/\kappa^n$, with higher powers more desirable due to the smallness of $\kappa$ compared to other energy scales. The deuteron electric polarizability and the deuteron mean-square charge radius in leading-order zero-range approximation  scale like $\frac{1}{\kappa^5}$ and $\frac{1}{\kappa^3}$, respectively, and have errors of roughly 3/4\% \cite{dipole,chi-pt,EFT-review} and 2\% \cite{rsq}, respectively. The leading {\sl fractional corrections} for both of these quantities are of order $(\kappa R_V)^3$. For these observables the corrections are small and the zero-range approximation clearly works very well \cite{d-pol}. Our leading-order nuclear polarization correction scales like $\frac{1}{\kappa^4}$, and we therefore expect results accurate to within 1-2\%. This has implications for the ultimate accuracy of polarization corrections, which we will discuss in Section (4).

We have chosen to ignore small corrections to the nuclear physics, and concentrate here on the global properties of the zero-range approximation and their implications. We ignore the following: (1) the effect of the proton-neutron mass difference, which increases the square of the dipole operator by $\sim 0.1\%$ \cite{ingo}; (2) meson-exchange-current contributions to the tiny  magnetic  sum rule, which are expected to be $\sim$ 15\% \cite{MEC} of the magnetic contribution, or $\sim 0.1\%$ of the dominant term; (3) meson-exchange (i.e., potential-dependent) contributions to the deuteron charge operator, which are of relativistic order \cite{dphoto} and thus should be $\sim$ 0.1 $-$ 0.2\% of the dominant term; (4) all other relativistic corrections to the nuclear physics, which are also expected to be $\sim$ 0.1~$-$~0.2\% \cite{dipole,chi-pt}; (5) phase shifts in all but s-waves. The corrections (4) will be discussed in some detail at the end of Appendix B, but will not be implemented in the numerical results. Potential models automatically include (5), but typically don't incorporate (1)-(4).

In this section we will combine our previously derived polarization corrections (expressed in terms of energy-weighted sum rules) with the zero-range approximation evaluation of those sum rules in Appendix B.  Our numerical results are listed in Table I and compared with the ``exact'' numerical results of Ref.~\cite{KP}, which used the Argonne V18 potential model \cite{AV18} to calculate nuclear matrix elements. Note that attractive (negative) contributions to the 2S energy will {\sl increase} the 2P-2S Lamb shift and thus are {\sl positive} entries in the table and in the individual entries $\Delta E_{\rm pol}$ listed below.

The  energy scale for deuteron polarization corrections in the 2S state is set by the prefactor in Eqn.~(11) evaluated for $Z=1$ and $n=2$:
$$
\frac{4 \pi}{3} \alpha^2 |\phi_n (0)|^2 = \frac{4 Z^3 \alpha^5 (m_r c^2)^3}{3 n^3 (\hbar c)^2} = 0.088636 \;\; {\rm meV\;  fm^{-2}} \, . \eqno(12)
$$
We note that all masses in this problem are known to sufficient accuracy that they do not impact the overall uncertainty. In addition each contribution  will be written as the prefactor in Eqn.~(12) times everything else. We will also insert factors of $\hbar$ and $c$ to produce results in accordance with SI units, and also equate $2 \mu$ to $M_N$. Errors and uncertainties should be judged on the scale of 1\% of the largest term or about 0.020 meV, which is $\sim 5 \epsilon_p$.

The leading-order result for the 2P-2S polarization correction from the first term in Eqn.~(11) plus Eqn.~(B16a) arises from the C1 multipole, which is driven by the unretarded dipole operator obtained from  $\rho_{\rm ch} (\vx)$:
$$
\Delta E_{\rm pol}^{\rm C1} = \left[ \frac{4 Z^3 \alpha^5 (m_r c^2)^3}{3 n^3 (\hbar c)^2} \right] \left [ \frac{4 \sqrt{2 m_r M_N c^2 }\,A_S^2}{35 \pi\, \kappa^{4}\, \hbar} \right ]  \, . \eqno(13)
$$
The quantity $A_S^2$ has an uncertainty of 0.2\%, which dominates Eqn.~(13). This is typical of all zero-range calculations. The first entry in Table 1 shows that our zero-range result (despite its incredibly simple form) differs by only 0.8\% from the numerical calculation in Ref.~\cite{KP} of the same quantity using the AV18 potential model. The $1/\kappa^4$ behavior is the reason for the accuracy, and the implications of this will be discussed in Section~(4).

\begin{table}[htb]
\centering
\caption{Contributions in meV to the 2P-2S $\mu$-d Lamb shift from the sum of the 2S polarization corrections and the 2S {\sl elastic} Zemach term. Except for nucleon-finite-size contributions (labelled f.s. and listed last) all nuclear polarization terms are labelled by their multipole (all are charge multipoles, CL, except for the magnetic dipole, M1), the equation number for that contribution is listed next, ZRA labels the zero-range approximation results of this work, the numerical results of Ref.~\cite{KP} are next, followed by their absolute difference and percentage difference. The running sum of contributions for the zero-range results and those of Ref.~\cite{KP} are given last. Blank ``f.s.'' entries result because Ref.~\cite{KP} assumed point nucleons, which eliminates the nucleon-finite-size contributions. The last entry sums the C0, retarded C1, and C2 multipoles listed in entries 3-5, which arose as a {\sl single} term in Eqn.~(6).
}
\hspace{0.25in}

\begin{tabular}{||r|c|rr|rr|cc||}
\hline
multipole     & Eqn. &ZRA & Ref.~\cite{KP}  & diff{  }  & \% &sum-0 &sum-\cite{KP}\\ \hline\hline\hline
       leading C1& 13  &1.925    & 1.910     &  0.015  & 0.8     & 1.925  & 1.910  \\ \hline
sub-leading C1& 14  & -0.037  & -0.035    & -0.002  & 7.0    & 1.888  & 1.875   \\ \hline \hline
                    C0& 16  & -0.042  & -0.045    & 0.003   & -7.6   & 1.846 & 1.830   \\ \hline
     retarded  C1& 17  &  0.137  & 0.151     & -0.014  & -9.4   & 1.983 & 1.981   \\ \hline
                    C2& 18  & -0.061  & -0.066    &  0.005  & -7.9    & 1.922 & 1.915    \\ \hline\hline
                    M1& 19  & -0.011  & -0.016    &  0.005  & -34.0 & 1.912 & 1.899   \\\hline\hline\hline
$\langle r^3 \rangle_{(2)}^{p p}$ f.s.& 15  &  0.030  &   &   & & 1.942 &      \\ \hline
 pn correl. f.s.   & 15  & -0.023  &               &             &          & 1.920 &      \\ \hline
retarded C1 f.s.& 17  &  0.021  &               &             &          & 1.941 &      \\ \hline\hline\hline
 C0+ret-C1+C2 &  & 0.034  & 0.040    &  -0.006  & -14.0 &  &    \\\hline
\end{tabular}
\end{table}

The next entry in Eqn.~(11) plus Eqn.~(B16b) form the sub-leading-order C1 term that arises as a relativistic correction in the atom
$$
\Delta E_{\rm pol}^{\rm sub-C1} = - \left[ \frac{4 Z^3 \alpha^5 (m_r c^2)^3}{3 n^3 (\hbar c)^2} \right] \left [ \frac{A_S^2\; \hbar}{5 \pi\, \kappa^{2}\, \sqrt{2 m_r M_N c^2}} \right ]  \, , \eqno(14)
$$
and is the second item in Table 1. Note the $1/\kappa^2$ behavior, which accounts for the lesser accuracy. Given the small size of the term, however,  that inaccuracy is not significant.

The next entry in Eqn.~(11) results from adding the  {\sl elastic} Zemach moment contribution to the {\sl inelastic} Zemach-like contribution. The inelastic term has mixed multipolarity and would be quite difficult to calculate. However, when  the corresponding elastic  part is added (as discussed below Eqn.~(7)) the summed result is a simple matrix element, is much easier to calculate, is less model dependent (see Eqn.~(B12)), and is much smaller. The review by Borie \cite{borie} lists a contribution from the deuteron elastic Zemach moment of 0.433 meV. In the limit of point nucleons this would be exactly canceled by the inelastic contribution. For extended nucleons there are two contributions that largely cancel and sum to .008 meV. The first is from the proton's third Zemach moment (seventh entry in Table 1), while the second arises from the Zemach moment due to the overlapping proton and neutron charge distributions. While the latter term is model dependent, over 80\% of its contribution (eighth entry in Table 1) arises from a model-independent operator (see discussion above Eqn.~(B12)). These two finite-size contributions are obtained in zero-range approximation by combining the third term in Eqn.~(11) with Eqns.~(B12), (B13), and (B14)
$$\eqalignno{
&\!\!\!\Delta E_{\rm pol}^{\rm fs} = \left[ \frac{4 Z^3 \alpha^5 (m_r c^2)^3}{3 n^3 (\hbar c)^2} \right] \times \cr
&\!\!\! \left [\frac{ m_r c \langle r^3 \rangle_{(2)}^{p p}}{4 \hbar} - \frac{3 m_r\, c\, \lambda\, A_S^2}{2 \hbar \kappa^2} + \frac{6 m_r \,c\, \lambda A_S^2}{\hbar \beta^2} \left(10 \ln{(2 \kappa/\beta)} + \frac{77}{12} + \cdots \right) 
 \right ] \, .  &(15) 
 }
$$
The first term in the bracket arises from the proton Zemach moment, the second term is from the model-independent part of the pn overlap contribution, and the third term is the small model-dependent part of the pn contribution that depends on the model parameter $\beta\gg \kappa$ that specifies the size of the nucleons. Numerical values for $\beta$, $\lambda$, and  $\langle r^3 \rangle_{(2)}^{p p}$ are given below Eqn.~(B11).

Our next three entries arose as a single $z^4$-term in Eqn.~(6), but were split into the separate contributions of three multipoles in Eqn.~(11): C0, retarded C1, and C2, which we will discuss in that order. 

We combine the fifth term in Eqn.~(11) with Eqn.~(B27) to obtain the monopole (i.e., C0) result, which is also the third entry in Table 1:
$$
\Delta E_{\rm pol}^{\rm C0} = -\left[ \frac{4 Z^3 \alpha^5 (m_r c^2)^3}{3 n^3 (\hbar c)^2} \right]   \left [\frac{A_S^2 \, }{6 \pi \hbar \kappa^{4} } \sqrt{ \frac{m_r^3 c^2}{2  M_N}} \,\left( \bar{a}^2 G_3 +2 \bar{a} \bar{b}\, G_4 +\bar{b}^2\, G_5 \right) \right ]\, . \eqno({\rm 16})
$$
One significant feature of the C0 sum rule is that it involves excitation of $^3S_1$ continuum waves, which have a large scattering length ($a_t \sim 5.4$ fm) that significantly modifies their asymptotic form. The modifications necessary to treat this are thoroughly discussed in Appendix B. In practical terms the effect of incorporating $a_t$ is to lower the C0 result by 17\%. The constants $\bar{a}$ and $\bar{b}$, and the functions $G_m$ (which depend only on $y = \kappa a_t$) are defined below Eqn.~(B23). We note that the quantity in parentheses in Eqn.~(16) involving the $G_m$ equals 106/315 for $y = 0$.

The retarded C1 result can be obtained by combining  the fourth term in Eqn.~(11) with Eqns.~(B18), (B19), and (B16b). It consists of two terms in Eqn.~(17): the point-nucleon contribution  and the contribution from the finite size of the neutron and proton
$$\eqalignno{
\Delta E_{\rm pol}^{\rm ret-C1} & = \left[ \frac{4 Z^3 \alpha^5 (m_r c^2)^3}{3 n^3 (\hbar c)^2} \right] 
\left [\frac{16 A_S^2 }{105 \pi \hbar \kappa^{4}} \sqrt{ \frac{m_r^3 c^2}{2  M_N}}\right ] \cr
& + \left[ \frac{4 Z^3 \alpha^5 (m_r c^2)^3}{3 n^3 (\hbar c)^2} \right] \left[ \frac{8\, A_S^2 }{15 \pi \hbar \kappa^{2}} \sqrt{ \frac{m_r^3 c^2}{2  M_N}} \left ( \langle r^2 \rangle_p - \langle r^2 \rangle \nr2 \right )\right] \, . &({\rm 17})
}
$$
The point-nucleon result is listed on the first line of Eqn.~(17) and the fourth line of  Table 1, while the nucleon finite-size contribution is listed on the second line of Eqn.~(17) and the ninth line of Table 1. Numerical values of the nucleon radii are listed below Eqn.~(B11). Note that the sum rule required for the last line of Eqn.~(17) (see Eqn.~(B19)) is the same one needed for Eqn.~(14). We can infer that sum rule from the numerical results of Ref.~\cite{KP}. That more accurate value lowers the zero-range result in line 9 of Table 1 from 0.021 meV  to 0.020 meV.

The quadrupole or C2 contribution is obtained by combining the last term in Eqn.~(11) with Eqn.~(B21)
$$
\Delta E_{\rm pol}^{\rm C2}  = -\left[ \frac{4 Z^3 \alpha^5 (m_r c^2)^3}{3 n^3 (\hbar c)^2} \right] \left [\frac{64 A_S^2 }{945 \pi \hbar \kappa^{4}} \sqrt{ \frac{m_r^3 c^2}{2  M_N}}\right ]  \, ,  \eqno({\rm 18})
$$
and is listed on line 5 of Table 1.

The last three contributions (items 3-5 in Table 1) are separately not negligible, although there is substantial cancellation and a relatively small net result. {\sl This cancellation is not accidental for point-like nucleons}. Sum rules with 1/2-integer energy weightings are more difficult to manipulate than those with integer weightings that dominate the electronic atom case. A  $z^4$ sum rule with a linear energy weighting ($\omega_N$ rather than $\sqrt{\omega_N}$ in this case)  is  shown in Eqn.~(B36) to vanish  in the zero-range approximation. Therefore our closely related sum rule (differing only by $\sqrt{\omega_N}$ rather than $\omega_N$) can be expected to be small {\sl in all realistic calculations} due to cancellations. This has implications for uncertainties that will be discussed in the next section.

Our final contribution arises from the magnetic dipole (viz., M1) interaction between the muon's current and the nuclear current, and is obtained by combining Eqns.~(A5), (A19), (A20), and (B34)
$$
\Delta E_{\rm pol}^{\rm M1}  = -\left[ \frac{4 Z^3 \alpha^5 (m_r c^2)^3}{3 n^3 (\hbar c)^2} \right] \left[\frac{A_S^2\, \mu_v^2\, \hbar^3}{\pi \sqrt{2 m_r M_N^5\, c^6}} \, \left((1-y)^2 G_1 (1-y^2) \right) \right] \,  ,\eqno({\rm 19})
$$
where $y=\kappa a_s$. The $^1S_0$ scattering length, $a_s$, is discussed above Eqn.~(B22) and the function $G_1$ is defined below Eqn.~(B23).   This  term involves only s-waves and requires special treatment because $|y|$ is so large. The sum rule for $y=0$ is actually logarithmically divergent. However, if we take the log-divergent number to be $\sim1$, the quantity in parentheses above is lowered by roughly 10\%. This change is small because the  limit for very large $|y|$ of the expression in the large parentheses in Eqn.~(19) is 1.
The numerical result for this magnetic contribution is listed in line 7 of Table 1. The huge factor of $M_N^{5/2}$ in the denominator is partially compensated by the factor of $\mu_v^2$, the square of the nucleon isovector magnetic moment ($\mu_v = \mu_p - \mu_n =  4.706\, \mu_N$ \cite{pdg}).

A variety of other contributions from the charge, current and seagull structure functions can be estimated and are much smaller than the M1 term calculated above and in Table 1. They are therefore negligible on the scale of the 0.016~meV uncertainty estimate of Ref.~\cite{KP}, and small on the scale of the $\epsilon_p \sim$ 0.004~meV uncertainty in the $\mu$-p experiment. The forms of these small terms are listed in the subsection {\sl Gauge Sum Rules} of Appendix B. The slow asymptotic behavior in $\omega_N$ of the current structure function introduces some uncertainty, however, which is discussed in the next section.

\section{Discussion and Conclusions}

Several pertinent remarks can be made based on Table 1. The first is that only one of the small contributions is as large as 7\% of the dominant term, and the rest are much smaller. The three largest corrections comprise the canceling group of three multipoles, whose sum is listed in the bottommost entry. These cancellations occur at roughly the same level in both the results of Ref.~\cite{KP} and in zero-range approximation, and in all likelihood occur for all realistic potential models (in addition to the AV18 model used by Ref.~\cite{KP}). If this behavior holds for the set of modern potential models with realistic pion-range forces and different short-range behaviors, then estimates of the uncertainty in these multipoles should be set by the size of their {\sl sum} and not by any individual element.

The smaller of the zero-range contributions in Table 1 all deviate more from those of Ref.~\cite{KP} than does the dominant term. The reason is that these sum rules involve higher powers of $\omega_N$ and will be more sensitive to details of the interior part of the wave functions because of oscillations in the continuum wave functions. The dominant zero-range contribution is larger than the potential-model result, which should be expected for sum rules that saturate at very low energies (N.B., the zero-range wave functions do not satisfy the finiteness boundary condition at the origin). Four of the five smaller contributions are smaller in magnitude, however. The tiny M1 contribution is particularly problematic, and this zero-range approximation is not very accurate. Note however that because of the rapid decrease of the size of the secondary contributions their {\sl absolute} errors are not large and are of order $\epsilon_p$. Due to cancellations the running sums for the six point-nucleon terms of our zero-range approximation (1.912 meV) and the full results of Ref.~\cite{KP} (1.899 meV)  are within the stated uncertainty (0.016 meV) of the latter work. This is somewhat remarkable given the minimal amount of physics needed for the zero-range calculation.

The uncertainty in each finite-size contribution should also be no larger than $\epsilon_p$. The proton Zemach term is measured and the uncertainty is less than 5\%. Most of the p-n correlation term is model-independent. The retarded C1 finite-size term depends on accurately (enough) measured nucleon sizes and on the same dipole sum rule that determines the sub-leading C1 contribution in line 2 of Table 1.

Based on how well our zero-range approximation tracks the AV18 calculation we believe that the uncertainty in each smaller contribution is likely not significantly larger than $\epsilon_p$. {\sl This should be checked by performing calculations with more potential models} that have quality fits to the scattering data, and thus agree with experimental values for $E_B$ and $A_S$ (an absolutely essential requirement). In this regard we note that the AV18 potential model \cite{AV18} has $A_S$ = 0.8850 fm$^{-1/2}$, which is slightly larger than the value of 0.8845 fm$^{-1/2}$ that we used and  would increase our dominant term by 0.002 meV, or $\epsilon_p/2$. Had we used this value with the zero-range approximation our dominant-term discrepancy with Ref.~\cite{KP} would have increased to 0.9\%.

It therefore seems likely that the uncertainty in the theoretical calculation of the set of polarization corrections discussed here is set by the uncertainty in the dominant term, where the zero-range  and AV18 results differ by slightly less than 1\%. {\sl A most informative  test of this assumption would be to use effective field theory} along the lines of Ref.~\cite{chi-pt}, in which interactions in the p-waves and relativistic corrections to the nuclear physics were systematically incorporated into the deuteron electric polarizability. The sum rules for the deuteron electric polarizability and the dominant $\mu$-d polarizability term are very similar (differing only by a factor of $\sqrt{\omega}$).

We emphasize that the only substantive difference between our results in Table (1) and those of Ref.~\cite{KP} are the small 0.029 meV nucleon-finite-size contributions. These are easily incorporated, although a better estimate than ours is both warranted and possible.

Reference~\cite{KP} also calculated the $Z\alpha$ and $(Z\alpha)^2$ Coulomb corrections to the polarizability, finding contributions to both the 2S and 2P levels. The corrections of order $(Z \alpha)^2$ and for the 2P levels in order $Z \alpha$ are new. The 2S-state $Z \alpha$ correction contains a constant term and a very large term proportional to $\ln{(2 m_r Z^2 \alpha^2/\omega_N)}$, both terms contained in a sum over dipole excitations weighted by $1/\omega_N$. These contributions had been previously calculated in Ref.~\cite{myhe4}. The logarithmic term is dominant and both calculations agree on its form, but the constant term for the 2S state in Ref.~\cite{myhe4} differs and is presumably in error.

We also note that there is a non-negligible contribution to the $\mu$-p \cite{carl,mohr,KP1,KP2,martyn} and $\mu$-d \cite{KP} polarizability corrections from the intrinsic electromagnetic polarizabilities of the proton. In the deuteron this should be supplemented by the intrinsic polarizability contributions of the neutron. The essential equality of the electromagnetic polarizabilities of the neutron and proton \cite{pdg} suggests that the neutron contribution is of comparable size to that of the proton. 

One potential problem must be resolved before any attempt is made to shrink the uncertainty in the polarization corrections. It was recently pointed out \cite{Ji,TRIUMF} that there is considerable high-energy strength in the electric dipole (i.e., E1) part of the {\sl current} structure function. This strength would lead to non-negligible higher-order energy-weighted dipole sum rules (corresponding to energy weightings of $\omega_N^{3/2 + n}$ with $n \ge 0$), and would indicate that some expansions in $\omega_N /m$ do not  converge very rapidly. These sum rules are divergent in zero-range approximation, and estimates of size are therefore problematic. Numerically integrating the slowly converging part of the E1 structure function in zero-range approximation gives an attractive contribution of 0.024 meV or roughly $6 \epsilon_p$, which is commensurate with the estimates of Ref.~\cite{Ji,TRIUMF}. Much of the strength results from nuclear energies above 200 MeV. Whether this problem exists in other partial waves is unknown. A discussion of this convergence problem and asymptotic properties in $\omega_N$ is provided in the subsection {\sl Asymptotic Properties} in Appendix A.

\section*{Acknowledgements}

The author would like to thank F. Kottmann, R. Pohl, and A. Antognini of PSI, R. B. Wiringa, K. Pachucki, I. Sick, E. Borie, S. Bacca, C. Ji, D. R. Phillips, and H. W. Griesshammer for very helpful discussions.

\section*{Appendix A - Nuclear Structure Functions}

In this appendix we derive the general structure functions that subsume the non-relativistic one derived in Eqns.~(2) and (5) in the main text, at least for the lower-order terms. The general structure functions do not incorporate recoil or reduced-mass effects, and $m$ below refers simply to the lepton mass.

\subsection*{Exact Structure Functions}

The energy shift due to  nuclear polarization for the n\underline{th} hydrogenic S-state in order $\alpha^5$ is most conveniently calculated by performing the  
contour integral over the time component of the virtual momentum $q^{\mu}$ (i.e., $q_0$) in the loops 
of Fig.\ (1) in Coulomb gauge. This was implemented in Ref.~\cite{roland} and leads to
$$\eqalignno{
 \Delta E_{\rm pol} 
&= -8 \alpha^2 m  
| \phi_n (0) |^2 \int \frac{d^3 q}{4 \pi} \left[ \sum_{N \neq 0} 
\left[ \frac{(2E + \omega_N ) | \langle N| \rho_{\rm ch} (\vq) |0 \rangle |^2}{E q^4
[(E + \omega_N)^2 - m^2]} \right. \right. \cr 
&+ \left. \left( \left[ \frac{q^2}{4 m^2} \right] 
\frac{2 E + \omega_N}{E q^4 [ ( E + \omega_N )^2 - m^2]}  
-\frac{(2 q + \omega_N)}{4 m^2 q^3 (q + \omega_N)^2}
\right) | \langle N | \vJ_{\bot} (\vq) | 0 \rangle |^2 \right] \cr 
&+  \left. \frac{B^{ii \bot}_{\rm in} (\vq)}{8 q^2 m^2}  
\left( \frac{1}{q} - \frac{1}{E} \right) \right], &({\rm A1}) 
}
$$    
where $q^2 \equiv \vq^2, E = \sqrt{q^2 + m^2}$, and $\omega_N =  
E_N - E_0$ is the energy of excitation (relative to the ground state) of  
the N\underline{th} nuclear state (which by assumption cannot be the  
ground state).  Unlike the charge terms both the current and seagull terms have infrared divergences, which cancel due to gauge invariance.

The nuclear physics is defined in terms of three nuclear operators: the nuclear charge operator that was introduced in Section~(2)
$$
\rho_{\rm ch}(\vq) = \int d^3 x \, e^{i \vq \cdot \vx}  \rho_{\rm ch} (\vx)\, , \eqno({\rm A2}) 
$$
the nuclear current operator
$$
\vJ (\vq) = \int d^3 x \, e^{i \vq \cdot \vx}  
\vJ (\vx)\, ,  \eqno({\rm A3})
$$
and the inelastic part of the nuclear seagull (two-photon) operator
$$
B^{ij}_{\rm in} (\vq) = \int d^3 x  \int d^3 y \, e^{i  
\vq \cdot (\vx - \vy)} B^{ij}_{\rm in} (\vx , \vy)\, . \eqno({\rm A4})
$$
We have used ``$\bot$'' to signify {\sl transverse}, or contraction with respect to ($\delta^{ij} -  
\hat{q}^{i} \hat{q}^{j}$). That is, $\vJ^{\, 2}_{\bot} \equiv J^i J^j  
(\delta^{ij} - \hat{q}^i \hat{q}^j)$.  This means that there is no longitudinal contribution from the current and seagull in Coulomb gauge.  Gauge invariance requires the longitudinal term to cancel the non-static part of the interaction between charges. This greatly simplifies the result since the charges then interact via a {\sl static} Coulomb force. Gauge invariance of the underlying {\sl inelastic} nuclear  Compton amplitude also restricts $B^{ij}$ to only the ``inelastic'' part, $B^{ij}_{\rm in}$ (discussed in some detail in Appendix B of Ref.~\cite{higher}). 

It is important to note that the partitioning of the nuclear Compton amplitude into ``inelastic'' contributions and seagull contributions is largely arbitrary. It depends entirely on what degrees of freedom in the problem are chosen to be {\bf active} and what are {\bf frozen}, which means they are not treated explicitly. One typically freezes higher-energy degrees of freedom, such as nucleon-antinucleon pairs and  intranuclear pions. The frozen ``pair''  degrees of freedom have an energy scale $\sim 2 M_N$ and generate the usual $e^2 \vA^2/2 M_N$ seagull term in the non-relativistic Schr\"odinger equation, since the Dirac equation with electromagnetic interactions has no seagulls at all. The best example of this is the neutron electric polarizability, which can be treated in several different ways \cite{review}, all of which lead to the same answer if calculated consistently. In a similar fashion freezing the pion degrees of freedom leads to nuclear potentials, meson-exchange currents, and seagull terms. If one starts with a gauge-invariant formalism a consistent treatment will result in one. A gauge-invariant definition of  observables will not change, but contributions to those observables can shift between inelastic and seagull terms \cite{review} depending on what degrees of freedom are frozen.

We proceed along the lines of Section~(2) and perform the $\vq$ integral using the definitions in Eqns.~(A2), (A3), and (A4). The exponentials in those equations can be collected into the form $e^{i \vq \cdot \vz}$, where $\vz = \vx-\vy$. Since the Lamb shift does not depend on the deuteron's azimuthal quantum numbers, the resulting integration must lead to scalar functions of $\vz$ for the charge contribution, and simple tensors for the transverse current and seagull terms. The integration is much more complicated than the NR derivation of Section~(2), but leads to results that are generically similar in form:
$$\eqalignno{
\Delta E_{\rm pol} &= -8 \alpha^2 m | \phi_n (0) |^2  
\int d^3 x \int d^3 y \left[ \sum_{N \neq 0}   \langle 0 | \rho_{\rm ch} (\vy) |N \rangle \langle N | \rho_{\rm ch} (\vx) |0 \rangle I_N (z)  \right. \cr
&+ \left.  \sum_{N \neq 0} \langle 0 | \vJ^i (\vy) |N \rangle \langle N | \vJ^j (\vx) |0 \rangle \left(\delta^{ij} J_N (z) + z^i z^j \bar{J}_N (z) \right) \right. \cr
&+ \left. \frac{1}{2} B^{ij}_{\rm in} (\vx , \vy) \left(\delta^{ij} K (z) +  z^i z^j \bar{K} (z) \right)\right] 
\, . &{\rm (A5)}
}
$$
All of our effort here will be devoted to obtaining the polarization  structure functions:  $I_N (z), J_N (z), \bar{J}_N (z), K (z)$, and  $\bar{K}(z)$.  After developing general forms we will perform  
appropriate tractable expansions. Although a similar calculation was performed in Ref.~\cite{higher} for the e-d atom, that approach must be modified because the electron mass was smaller than any $\omega_N$, which clearly does not hold for the muon mass. This mismatch in energy scales in polarization corrections means that the low-mass electron is driven into a relativistic regime. 
The much heavier muon is largely non-relativistic, and the required expansions are therefore quite different.

We begin with the dominant term, $I_N (z)$, which determines the interaction between charges and is  
the most difficult to obtain.  All other integrals can be obtained from  $I_N (z)$:
$$\eqalignno{
I_N (z) &= \frac{1}{\omega_N z}  
\int^{\infty}_{0}\frac{dq \, \left(2 E\, \omega_N + \omega_{N}^2 \right)\, \sin
(qz)}{q^3 E[(E + \omega_N)^2 - m^2]} \cr
&= \frac{1}{\omega_N z} \int^{\infty}_{0} \frac{dq\, \sin (qz)}{q^3  
E} \left[ 1 - \frac{q^2}{(\omega_N + E)^2 - m^2} \right] \cr
&\equiv \frac{(I_0 (z) - I_1 (z) )}{\omega_N z}\, .
& {\rm (A6)}
}
$$
We added and subtracted $q^2$ in the parentheses in the top line, making the first part equal to the bracketed term in the denominator. This conveniently splits the integral into a nominally infrared-divergent integral $I_0 (z)$ that is independent of the state $|N\rangle$ plus a complicated but well-behaved part, $I_1 (z)$. As we found in Section~(2), nominally divergent terms that are constants cannot excite the nucleus  and can be discarded. Two derivatives of $I_0 (z)$ yield a tractable integral, and two integrals yield the final form
$$
I_0 (z) = - \frac{1}{2 m^3} \int^{\beta}_{0} d \beta^{\prime} (\beta -  
\beta^{\prime})^2 \; K_0 (\beta^{\prime})\, , \eqno({\rm A7})
$$
where $\beta = m\, z$ and $K_0 (z)$ is the usual modified Bessel function of order zero. 

We can perform a partial fractions expansion on the bracket in $I_1$
$$
I_1 (z) =  \int^{\infty}_{0} \frac{dq\, \sin (qz)}{q  E} \left[\frac{1}{(\omega_N + E)^2 - m^2} \right] \, , \eqno{\rm (A8)}
$$
which allows us to rewrite $ I_N (z)$ in a much more useful form
$$
 I_N (z) = \frac{ \bar{I}_N (\xi;z) -  \bar{I}_N (\xi^{\prime};z)}{2 m\, \omega_N\, z}  \, , \eqno{\rm (A9)}
$$
where we have defined $\xi = \omega_N +m$ and $\xi^{\prime} = \omega_N - m$ together with
$$
\bar{I}_N (\xi;z) = \xi I_0 (z) + \int^{\infty}_{0} \frac{dq\, \sin (qz)}{q  E (E +\xi)} \, . \eqno{\rm (A10)}
$$
The denominator in the integral  in Eqn.~(A10) cannot vanish, but functions can {\sl smoothly} change form at $\xi^{\prime}/m$ = 0 and 1 (i.e., $\omega_N/m$ = 1 and 2). The trick used in Appendix A of Ref.~\cite{higher} to evaluate the integral works for all $\xi$, but only for $\xi^{\prime} > 0$, which was adequate for the e-d atom but not for our $\mu$-d atom case.  For the case involving $\xi^{\prime} < 0$ we note that a redefinition of parameters allows a trick similar to Eqn.~(A13) of Ref.~\cite{higher} to be used. The calculation is  long and tedious, and we simply quote the results.

We define parameters $w = \omega_N/m > 0$, $\mu^2 = \xi^2/m^2 -1 = 2 w + w^2 > 0$, $\mu^{\prime\, 2} = \xi^{\prime\, 2}/m^2 -1 = - 2 w + w^2$ [$>0$ for $w>2$], and $\nu^{\prime \, 2} = - \mu^{\prime\, 2} =  2 w - w^2$ [$>0$ for $w<2$]. We then obtain from Ref.~\cite{higher}
$$\eqalignno{
&\bar{I}_N (\xi ; \beta) = -\frac{\pi}{2 m^2 \mu^2} ( 1 - \cos (\mu \beta))  
+ \frac{\sin (\mu \beta) \sinh^{-1} (\mu)}{m^2 \mu^2}  \cr
&  - \frac{\xi}{m^3 \mu^2} \int^{\beta}_0 d \beta^{\prime} K_0 (\beta^{\prime}) \left( \cos (\mu (\beta- \beta^{\prime}))  
-  1  + \frac{\mu^{2}}{2} (\beta- \beta^{\prime})^2  \right)  \, . &({\rm A11})
}
$$
Note that the function $I_0$ in Eqn.~(A10) has become the final subtraction term in the integral in Eqn.~(A11). This compact expression composed of three terms can be expanded in powers of $\beta$ to provide tractable forms for computation. Taking into account the $1/z$ factor in Eqn.~(A9) we note that the first term above generates only {\sl odd} powers of $z$ in $I_N (z)$, while the remaining two terms generate only {\sl even} powers (including logarithms that begin with $z^4$). All Zemach-like terms (odd powers of $z$) therefore arise only from the first term. The leading power in that expansion does not involve the parameter $\mu$ and will cancel between the two $\bar{I}_N$ functions in Eqn.~(A9), leaving Zemach-like terms of order $z^3$, $z^5$, etc. The integral (i.e., third) term generates a leading-order $z^4$ power that contains a $\log{\beta}$ factor. The second ($\sim \sin{\mu \beta}$) term generates the smallest power of $z$ (viz., $z^2 = x^2 + y^2 -2 \vx \cdot \vy$), which produces the dominant dipole contribution in Eqn.~(A2), as shown in detail below Eqn.~(6) in Section~(2).

We note that for our problem the parameter $w = \omega_N/m$ is typically very small because the deuteron binding energy sets the scale for the important range  of excitation energies, and therefore $\mu^2$ is small. Expanding $\bar{I}_N (\xi ; \beta)$ to order $(\mu \beta)^5$ we find
$$\eqalignno{
\bar{I}_N (\xi ; \beta) \cong & \frac{-\pi z^2}{4}\left(1-\frac{\mu^2 \beta^2}{12}  + \cdots \right) 
+\frac{z \sinh^{-1}{\mu}}{m \mu } \left(1-\frac{\mu^2 \beta^2}{6} + \frac{\mu^4 \beta^4}{120} + \cdots \right) \cr
&+ \frac{\xi m^2 \mu^2 z^5}{120} \left(\gamma + \ln{(\beta/2)} -\frac{137}{60} + \cdots \right)\, , & (\rm{ A12})
}
$$
which is relatively simple. 

The remaining term depends on $\xi^{\prime}$, and there are three energy regimes. For $\omega_N \geq 2 m$ one simply substitutes $\mu^{\prime}$ for $\mu$ and $\xi^{\prime}$ for $\xi$ in Eqns.~(A11) and (A12). This corresponds to the electronic atom case and does not interest us here. The regime that does interests us is $0 < \omega_N \leq m$, where $\nu^{\prime\, 2}$ is positive and leads to
$$\eqalignno{
&\bar{I}_N (\xi^{\prime} ; \beta) = \frac{\pi}{2 m^2 \nu^{\prime\, 2}} ( 1 - \cosh (\nu^{\prime} \beta))  
+ \frac{\sinh (\nu^{\prime} \beta)(\pi - \sin^{-1} (\nu^{\prime})) }{m^2 \nu^{\prime\, 2}}  \cr
& +  \frac{\xi^{\prime}}{m^3 \nu^{\prime\, 2}} \int^{\beta}_0 d \beta^{\prime} K_0 (\beta^{\prime}) \left( \cosh (\nu^{\prime} (\beta- \beta^{\prime})) -  1  - \frac{\nu^{\prime\, 2}}{2} (\beta - \beta^{\prime})^2  \right)  \, . &(A13)
}
$$
Most of the change in form between (A11) and (A13) is trivial because $\nu^{\prime\, 2} = - \mu^{\prime\, 2}$. With decreasing $w$ the parameter $\mu^{\prime}$ becomes imaginary; this  interchanges the roles of the trigonometric and hyperbolic functions and $\mu^{\prime\, 2}$ is replaced by $\nu^{\prime\, 2}$.  We note that in the third regime ($m \leq \omega_N \leq 2 m$) we simply replace $(\pi - \sin^{-1} (\nu^{\prime}))$ by $\sin^{-1} (\nu^{\prime})$.

Expanding $\bar{I}_N (\xi^{\prime} ; \beta)$ in Eqn.~(A13) to order $(\nu^{\prime} \beta)^5$ we find
$$\eqalignno{
\bar{I}_N (\xi^{\prime} ; \beta) &\cong \frac{-\pi z^2}{4}\left(1+\frac{\nu^{\prime\, 2} \beta^2}{12} + \cdots \right) \cr
&+\frac{z (\pi - \sin^{-1} (\nu^{\prime}))}{m \nu^{\prime }} \left(1+\frac{\nu^{\prime\, 2} \beta^2}{6} + \frac{\nu^{\prime\, 4} \beta^4}{120}+ \cdots \right) \cr
&+ \frac{\xi^{\prime} m^2 \nu^{\prime\, 2} z^5}{120} \left(\gamma + \ln{(\beta/2)} -\frac{137}{60} + \cdots \right)\, . & (\rm{ A14})
}
$$
Completing Eqn.~(A9) and dropping a constant term that cannot excite the nucleus produces
$$\eqalignno{
I_N (z) &\cong-\frac{\pi z^2 \nu^{\prime}}{12\, \omega_N} -\frac{z^2 a_2}{12 \, \omega_N} + \frac{\pi z^3}{24} -\frac{\pi z^4 m^2 \nu^{\prime \, 3}}{240 \omega_N} \cr
&+ \frac{m^2 z^4 a_4}{240 \omega_N} + \frac{z^4 \omega_N (\ln{(\beta/2)} + \gamma -137/60)}{40} \, , &({\rm A15})
}
$$
where $a_2 = \mu \sinh^{-1} (\mu) -\nu^{\prime} \sin^{-1} (\nu^{\prime})$ and $a_4 = \mu^3 \sinh^{-1} (\mu) +\nu^{\prime\, 3} \sin^{-1} (\nu^{\prime})$. Equation (A15) is correct to order ($z^4$), but an expansion of $\mu$ and $\nu^{\prime}$ in terms of the small parameter $w = \omega_N/m \sim 1/20$ is warranted in order to obtain tractable expressions. It is easy to show that both $a_2$ and $a_4$ are even functions of $w$ and we therefore need only the leading terms. We find that $a_2 \cong 2 w^2/3$, $a_4 \cong 8 w^2$, $\nu^{\prime}/w \cong \sqrt{2/w}(1 -w/4 -w^2/32 + \cdots)$, and $\nu^{\prime\, 3}/w \cong 2 \sqrt{2w}$, which produces our final result
$$\eqalignno{
I_N (z) &\cong -\frac{\pi z^2}{12\, m}\sqrt{\frac{2 m}{\omega_N}}\left(1-\frac{\omega_N}{4 m}\right)  + \frac{\pi z^3}{24} -\frac{\pi z^4 m}{60}\sqrt{\frac{\omega_N}{2 m}}  \cr
&-\frac{z^2 \omega_N}{18 \,m^2}+ \frac{z^4 \omega_N (\ln{(\beta/2)} + \gamma -57/60)}{40} + \cdots \, . &({\rm A16})
}
$$
We note that the three terms in the top line containing a factor of $\pi$ were derived by Ref.~\cite{KP} and the first term also by Ref.~\cite{myhe4}, while the first term in the second line was also found by Ref.~\cite{Ji}. The top line contains the non-relativistic terms developed in our Eqn.~(6) plus the relativistic correction derived below that equation. The two remaining terms proportional to $\omega_N$ are of marginal size and similar to terms from the currents and seagull. They will be considered together in Appendix B. Because the coefficients of the $z^2$ terms are crucial and change form in each regime of $w$, we verified  that these coefficients were equal to those obtained by numerically integrating the part of Eqn.~(A6) corresponding to that $z^2$ term.

Exact current and seagull structure functions $J_N (z), \bar{J}_N (z), K (z)$, and  $\bar{K}(z)$ can be obtained from our previous results using the tricks in Ref.~\cite{higher}. Because the scales in the current and seagull terms are intrinsically small, there is no need to display exact results, and leading terms in an expansion suffice. The functions $ K (z)$, and  $\bar{K}(z)$ were displayed in Eqns.~(A23) and (A24) of Ref.~\cite{higher}:
$$
\bar{K} \cong - \frac{1}{120} (\gamma + \log (\beta / 2) - 31/30) 
+ O (z^2)\, , \eqno ({\rm A17})
$$
and
$$
K \cong - \frac{\log ( 2 \lambda / m)}{6m^2}  -2 z^2 (\bar{K} + \frac{1}{480})
\, . \eqno ({\rm A18}) 
$$
Note the infrared cutoff, $\lambda$. 

The current structure functions $J_N (z)$ and $\bar{J}_N (z)$ can be directly obtained using the trick discussed just below Eqn.~(A18) of Ref.~\cite{higher}. We simply quote the results:
$$
\bar{J}_N (z) = - \frac{\pi}{30 m} \sqrt{\frac{\omega_N}{2 m}} + \frac{\omega_N}{20 m^2} \left( \ln{\left[\frac{m}{2 \omega_N}\right]} + 1\right) \, , \eqno ({\rm A19})
$$
and
$$
J_N (z) =  \frac{\pi}{12\, m^3} \sqrt{\frac{2 m}{\omega_N}}\left(1- \frac{\omega_N}{4 m}\right) + \frac{\omega_N}{18\, m^4} - \frac{\left(1 +\ln{\left(\frac{\omega_N}{\lambda}\right)}\right)}{6\, \omega_N\, m^2} -2 z^2 \bar{J}_N (z) \, . \eqno ({\rm A20})
$$
Note the infrared cutoff, $\lambda$.

The sum rules based on $J_N (z)$ and $\bar{J}_N (z)$ will be of two types. The terms involving $\bar{J}_N (z)$ have two powers of $\vz$ and correspond to M1, E0, retarded E1, and E2 excitations. They are all small except for the M1 case, where the square of the nucleon isovector magnetic moment is a factor of more than 20. We will evaluate that case in Appendix B. The second type involves the terms in $J_N $ that are independent of $z$. In that case Siegert's theorem \cite{higher} can be used in the form $ \langle N | \int d^3 x \, \vJ (\vx ) | 0 \rangle = i \omega_N \langle N | \vD | 0 \rangle$, which produces two additional powers of $\omega_N$. This generates sum rules with powers of $\omega_N$ that are $\geq 3/2$, which are divergent in zero-range approximation. These terms, moreover, have strength at large nuclear energies, and they are discussed together in the the subsection {\sl Asymptotic Properties} below. All sum rules with a single power of $\omega_N$ have special properties and will be treated together  in the subsection {\sl Gauge Sum Rules} in Appendix B.

\subsection*{Breit Approximation}

The exact charge structure function $I_N(z)$ is given in Eqn.~(A6) by
$$
I_N (z) = \frac{1}{z}  
\int^{\infty}_{0}\frac{dq \,}{q^3} \sin(qz)\left [\frac{2 E\,  + \omega_{N}}{E((E + \omega_N)^2 - m^2)} \right]\, , \eqno({\rm A21})
$$
and results from integrating the $q_0$ variable in the charge-interaction parts of Figs.~(1a) and (1b). The square bracket is the sum of the two diagrams, which can be reseparated into their two components, (1a) and (1b), respectively, as
$$
\frac{2 E\,  + \omega_N}{E((E + \omega_N)^2 - m^2)} = \frac{1}{2 m E} \left [ \frac{E+m}{E +\omega_N - m} - \frac{E-m}{E+\omega_N +m} \right] \, . \eqno({\rm A22})
$$
The first term is from Fig.~(1a) and has an energy denominator that is simply the difference in energies of the intermediate state ($E+\omega_N$) and the ground state ($m$) of the coupled lepton-nucleus system. The second term from Fig.~(1b) does not have this form, because it reflects lepton ``pair'' intermediate states. The energies of the  lepton-plus-nucleus intermediate  and ground states  in the denominator have the {\sl same} signs. This means that an interpretation is not possible  in terms of a conventional Hamiltonian that is the {\sl sum} of two parts for a system composed of two parts.

In 1929 Breit \cite{gregory} constructed a tractable Hamiltonian for two interacting relativistic electrons by summing their Dirac components together with a Coulomb potential between them. Although this Breit equation is not an exact representation of the physics, it has nevertheless proven very useful. One way to derive the Breit result is to modify the boundary conditions \cite{breit} for two interacting systems. One can equivalently change the sign of $\omega_N$ in the second term in Eqn.~(A22). We will do this by adding the sign-changed term to the first term to generate the Breit term (labelled BR), and then  subtracting the same sign-changed term from the second term to form a non-Breit correction term (labelled NB). The resulting expression is still exact, but neglecting the NB term results in the Breit approximation.

This manipulation then produces two terms, Breit plus non-Breit:
$$\eqalignno{
\left[ \frac{2 E\,  + \omega_N}{E((E + \omega_N)^2 - m^2)}\right] &= \left[ \left[ \frac{\lambda^{\prime\, 2}}{m \omega_N}\right]  \frac{1}{q^2 +\lambda^{\prime\, 2}}\right]_{\rm BR} \cr
 &+\left [  \left(\frac{1}{ m E}\right) \frac{\omega_N  (E-m)}{(E + m)^2 -\omega_N^2}\right]_{\rm NB} \, , &({\rm A23})
}
$$
where $\lambda^{\prime \, 2} = 2 m \omega_N (1 - \omega_N/2 m)$ was introduced below Eqn.~(6). Note that $\lambda^{\prime \, 2}/m^2 \equiv  \nu^{\prime\, 2}$, which we introduced above Eqn.~(A11). We see that this exact division of our charge structure function, $I_N (z)$, produces one term that is a trivial modification of our very simple non-relativistic result plus a complicated correction term that has {\sl at least} one power of $\omega_N$. The Breit term clearly becomes problematic as $\omega_N \rightarrow 2 m$, and Ref.~\cite{breit} contains a discussion of some of the diseases associated with this approximation.

\subsection*{Asymptotic Properties}

We have implicitly assumed that our sum rules saturate at low excitation energies. Reference~\cite{Ji} has nevertheless identified a part of the transverse E1 (i.e., unretarded dipole) polarization correction that has strength at very high nuclear excitation energies. The dipole parts of the unretarded charge structure function are those proportional to $z^2$ in an expansion of $I_N (z)$, while the corresponding parts of the transverse structure function are the $z$-independent parts of $J_N (z)$ and $K (z)$. 
Collecting terms we find that the complete unretarded dipole contribution is given by
$$\eqalignno{
&\Delta E_{\rm pol} = -\frac{4 \pi }{3}  \alpha^2 | \phi_n (0) |^2  \sum_{N \neq 0} | \langle N| \vD | 0 \rangle |^2 \; \times \cr
&\frac{1}{\pi}\left[ 2 b(w)  + \left( w^2 b(w) - w (1 + \ln{(\omega_N/\lambda)}) \right) + w \ln{(m/2 \lambda)} \right]\, , & ({\rm A24})
}
$$
where  $w = \omega_N/m$ and the dipole charge function $b(w)$ is defined as
$$\eqalignno{
b(w) =& \frac{1}{2 w} \left[\mu \sinh^{-1} (\mu) - \mu^{\prime} \sinh^{-1} (\mu^{\prime}) \right] \qquad \qquad [w \geq 2] \cr
=& \frac{1}{2 w} \left[\mu \sinh^{-1} (\mu) + \nu^{\prime} \sin^{-1} (\nu^{\prime}) \right] \qquad\qquad [2 \geq w \geq 1] \cr
=& \frac{1}{2 w} \left[\mu \sinh^{-1} (\mu) + \nu^{\prime} (\pi - \sin^{-1} (\nu^{\prime})) \right] \qquad [ 1 \geq w ] \, . & ({\rm A25})
}
$$
The parameters $\mu$, $\mu^{\prime}$, and $\nu^{\prime}$ were defined above Eqn.~(A11). In Eqn.~(A24) the first term in the square bracket is the charge contribution, the second term (in large parentheses) is the transverse current contribution, while the remaining term is  the seagull contribution. 

Note that the infrared cutoff ($\lambda$) in Eqn.~(A24) cancels, and that the final two transverse terms sum to $- w (1 + \ln{(2 w)})$. In addition we see a transverse current term ($\sim w^2 \, b(w)$) that is potentially more sensitive to high excitation energies than is the charge term. The tripartite definition of $b(w)$ in Eqn.~(A25) creates a complication, however,  in determining this sensitivity. Our expansions of the dipole structure function above and in the main text were based entirely on the form of $b(w)$ for $w \leq 1$ or $\omega_N \leq m_r \cong$ 100 MeV, and not on the highest-energy form corresponding to $w \geq 2$. The reason for this was the assumption that low excitation energies saturate the sum rules. Indeed, the factor of $ \nu^{\prime} \pi$ in the bottom line of Eqn.~(A25) is entirely responsible for all of the half-integral energy sum rules that we developed in Eqn.~(11). The $\sinh^{-1}$ and $\sin^{-1}$ terms play no role in those sum rule terms. We therefore can assume either that low-energy excitations saturate Eqn.~(A24) and expand the $w \leq 1$ form of $b(w)$ to check for high-energy sensitivity, or we can perform a true asymptotic expansion using the $w \geq 2$ form. Neither choice is entirely representative, so we will do both.

For $w \leq 1$ the bracketed term  in Eqn.~(A24) (including the prefactor of $1/\pi$) can be expanded as a series in $w$
$$
\left [ \sqrt{\frac{2}{w}} \left( 1 - \frac{w}{4} - \frac{w^2}{32} \right) + \frac{2 w}{3 \pi} \right ] + \left[ \sqrt{\frac{w^3}{2}} - \frac{w (1 + \ln{(2w)} )}{\pi} \right]  + {\cal{O}} (w^{5/2})\, . \eqno({\rm A26})
$$
The first bracket contains the charge contribution, while the second bracket contains the transverse terms. The terms with powers of $w$ that are $\leq 1$ were also obtained in a similar manner by Ref.~\cite{Ji}. Terms with powers larger than this diverge in zero-range approximation. Because of Siegert's Theorem the current terms are weighted more heavily towards higher energies. 

A very different result is obtained if one expands the $w \geq 2$ form for large $w$
$$
b(w) \rightarrow \frac{1 + \ln{(2w)}}{w} + {\cal{O}} (1/w^3) \, . \eqno({\rm A27})
$$
This guarantees that the charge and transverse contributions both behave asymptotically $\sim 1/w$, but a cancellation of large terms is required for the transverse result. The current spectral function therefore has the potential to be sensitive to high virtual excitation energies, unlike the charge spectral function. This can be checked for the E1 excitations by numerically integrating Eqn.~(A24), as was done in Ref.~\cite{Ji}. In zero-range approximation we find that the sum rules with powers of $w$ that are $\leq 1$ in the charge spectral function largely saturate at energies less than 100 MeV. Moreover, their sum agrees well with the numerically integrated one. There is no problem with the behavior of the charge spectral function. 

The transverse spectral function, however, presents a problem. Because of cancellations caused by the logarithm the $w$-linear term is  determined largely by virtual excitation energies larger than 200 MeV. This contribution can in principle be calculated using the closure trick introduced in Eqn.~(B35), and this should provide adequate accuracy for what is a rather small term. The $w^2\, b(w)$ term in Eqn.~(A24), however, provided none of the contributions to our final result, but  integrated numerically contributes an attractive 0.024 meV to the polarization correction in zero-range approximation, half of which comes from energies above 200 MeV.

This is a fairly serious problem because most potential models were not designed to be accurate at those energies. Moreover, it raises questions about the convergence of our procedure. Whether this problem exists for other transverse multipoles is unknown. Whether it is more or less severe when higher multipoles are summed is also unknown. This is a problem that needs to be resolved if polarization corrections with sub-1\% uncertainties are ever to be obtained.

\section*{Appendix B- Zero-Range Approximation}

\subsection*{Introduction}

In 1935 Bethe and Peierls \cite{zero} developed the zero-range approximation for the deuteron, which circumvented the almost complete lack of knowledge at that time about detailed properties of the force between the proton and neutron. It was known that the nuclear force had a short range ($R_V \sim 1$ fm) compared to the spatial extent of the weakly bound deuteron ($E_B \sim 2.2 $ MeV). They assumed that the range of the force could be neglected in many applications, and that only knowledge of the wave function of the deuteron {\sl outside the nuclear potential} was required for calculating many deuteron properties. This method has proven extremely useful in studies of deuteron photo-disintegration \cite{zero}, polarization corrections in the e-d Lamb shift \cite{rsq,don} and hyperfine splittings \cite{hyper,khrip}, the deuteron electric polarizabilities \cite{dipole}, and the deuteron charge radius \cite{rsq}. Its primary utility is that it can give a very simple and rather accurate estimate of some deuteron observables, and these estimates can  be systematically improved by incorporating more physics \cite{elpol}. There is a very substantial overlap between the zero-range approximation and some effective-field-theory treatments \cite{chi-pt,EFT-review} of the deuteron.

We will develop the simplest versions of this approximation (see Refs.~\cite{elpol,yulik} for improvement methods), and will use natural units ($\hbar = c =1$). Only non-relativistic dynamics will be considered until the final section of this appendix. Relativistic corrections were considered in Ref.~\cite{dipole}, and for the electric polarizability are expected to be $\lsim 0.1\%$. We use the conventional definition of $r$ as the distance between the proton and neutron. Then the non-relativistic Schr\"odinger equation for a bound state can be easily solved for the dominant s-wave in the absence of a potential or in the region outside a short-range potential
$$
\psi_S (r) = \frac{A_S}{\sqrt{4 \pi}} \frac{e^{-\kappa r}}{r} \, , \qquad \qquad [r > R_V] \eqno({\rm B}1)
$$
where $A_S = 0.8845(8)\, {\rm fm}^{-1/2}$~\cite{deut} is the experimental deuteron s-wave asymptotic normalization constant, and $\kappa = \sqrt{2 \mu E_B}$ = 45.7022 MeV is the deuteron ground-state virtual momentum (corresponding to 0.23161 ${\rm fm}^{-1}$ after dividing by $\hbar c$). The latter quantity is determined by twice the n-p reduced mass, $2 \mu =  938.918$ MeV, and the deuteron binding energy, $E_B = 2.224575(9)$ MeV \cite{deut}. Since $2 \mu$ is very close to $M_N$ (the average nucleon mass), the small dimensionless quantity $\kappa/M_N \cong 0.05$ is a relevant (and small) expansion  parameter.

The wave function in Eqn.~(B1) is clearly incomplete for $r < R_V$ and does not satisfy the finiteness boundary condition at the origin. We can produce one estimate of this error by computing the normalization
$$
\langle \psi_S | \psi_S \rangle = \frac{A_S^2}{2 \kappa} = \frac{1}{1-\kappa \rho_d} \, , \eqno({\rm  B}2)
$$
which follows from the definition $A_S^2 = \frac{2 \kappa}{1-\kappa \rho_d}$, where the deuteron effective range is $\rho_d$ = 1.765(4) fm \cite{deut}. The ``normalization'' of $|\psi_S|^2$ therefore equals 1.69 rather than 1, which is an overestimate of nearly 70\%. Of what quantitative use is a technique that is subject to such a large error? The key ingredient in the zero-range approximation is the smallness of $\kappa$ compared to other relevant deuteron energy scales, and this comparison improves for decreasing deuteron binding or for matrix elements containing more powers of $r$. In addition many corrections to the zero-range approximation scale as $(\kappa R_V \sim 1/4)^n$ for $n > 2$, and larger $n$ substantially improves the accuracy of the zero-range approximation. Equation (B2) is the worst case, and does not impact practical calculations.

\subsection*{Ground-State Radial Matrix Elements}

Matrix elements of positive powers of $r$  suppress the incorrect interior part of $\psi_S$ for $r < R_V$, while enhancing the correct exterior part. The error of the approximation therefore will dramatically decrease in such cases. Positive powers of $r$ will lead to matrix elements that depend on higher {\sl inverse} powers of $\kappa$, with the highest inverse powers being the most accurate. In addition the angular momentum barrier in more complicated observables that involve virtual excitations to non-s states will also suppress the interior part of the deuteron wave function and lead to higher inverse powers of $\kappa$. The deuteron electric polarizability and the deuteron mean-square charge radius in zero-range approximation scale like $\frac{1}{\kappa^5}$ and $\frac{1}{\kappa^3}$, respectively, and have errors of roughly 3/4\% \cite{dipole,chi-pt} and 2\% \cite{rsq}, respectively. We will see below that the leading term in the $\mu$-d polarization correction scales like $\frac{1}{\kappa^4}$ and has an error of slightly less than 1\%. These accuracies are sufficient to be quite useful, and the simplicity and relative accuracy of the zero-range results can lead to considerable insight about the importance of details of the nuclear force in a given calculation. We will see that the zero-range expansion is given in terms of simple observables, which are common features of {\sl all} quantitatively accurate nuclear force models.

The lack of an angular momentum barrier makes the s-wave virtual-excitation case special, because the interior region of the wave function becomes relatively more important. These cases therefore merit closer examination. The total charge operator is super-conserved and cannot cause transitions, so we will ignore matrix elements of this operator between the deuteron and its $^3S_1$ excited states. All other charge matrix elements involve powers of $r$, and they can and will be treated  in zero-range approximation. Ground-state matrix elements of constants will be treated exactly, yielding just those constants. This leaves only magnetic (viz., spin-flip M1) excitations as possible special cases, since they don't involve powers of $r$. These contributions will nevertheless be estimated in zero-range approximation, because they are quite small and great accuracy is not required.

\subsubsection*{Charge Operator Multipoles}

The ground-state matrix element of $r^n$ in zero-range approximation is given by
$$
\langle 0 | r^n | 0 \rangle\zrlow \equiv \langle r^n  \rangle \zrlow= A_S^2 \int_0^{\infty} d r\, r^n e^{-2 \kappa r} = \frac{A_S^2 \, n!}{(2 \kappa)^{n+1}}\, .  \eqno({\rm  B}3)
$$
This is typically not an observable, however. The nuclear charge operator in non-relativistic impulse approximation (no meson currents and no spin-orbit charge density, both of which are corrections of relativistic order) is given by
$$
\rho_{\rm ch} (\vx ) = \sum_{i = 1}^A \hat{\rho}_i (| \vx - \vx_i| )\, =  \sum_{i = 1}^A \hat{p}_i \; \rho_p (|\vx - \vx_i| ) + \hat{n}_i\; 
\rho_n (|\vx - \vx_i| )\, . \eqno({\rm  B}4)
$$
In this expression $\hat{\rho}_i (| \vx - \vx_i |)$ is the charge density at the point $\vx$ of nucleon $i$ expressed in terms of its position  $\vx_i$ {\sl relative to the nuclear CM}. This is further broken down into separate proton and neutron contributions, each with its respective isospin projection operator (viz., $\hat{p}_i = \frac{1 + \tau_z (i)}{2}$ and $\hat{n}_i = \frac{1 - \tau_z (i)}{2}$) and respective charge density (viz., $\rho_p (y)$ and $\rho_n (y)$). We note that $\rho_p (y)$ is normalized to 1, while $\rho_n (y)$ is normalized to 0. For the deuteron we ignore the small mass difference of the proton and neutron and use $\vx_1 = \vr/2$ and $\vx_2 = - \vr/2$.

Because the lepton in a hydrogenic atom carries small momentum compared to real or virtual momentum scales in nuclei, electromagnetic excitation of the lowest unretarded nuclear multipoles will dominate. We will require the following multipole charge operators:
$$
\int d^3 x \, \rho_{\rm ch} (\vx ) =  \sum_{i = 1}^A \hat{p}_i  =Z \, , \eqno({\rm  B5a})
$$ 
where Z is the number of protons in a nucleus with $A$ nucleons and $N$ neutrons. Note that we have chosen to define nuclear charges in multiples of the fundamental charge, $|e|$. The dipole operator is then given by
$$
\vD = \int d^3 x \; \vx\, \rho_{\rm ch} (\vx ) = \sum_{i = 1}^A \hat{p}_i \; \vx_i  = \sum_{i = 1}^A \frac{\tau_z (i)}{2} \; \vx_i  \rightarrow \frac{\vr}{2} \left ( \frac{\tau_z (1) -\tau_z (2)}{2} \right ) \, , \eqno({\rm  B5b})
$$ 
where the arrow points to the deuteron result. The mean-square radius operator is
$$
\hat{r}^2 = \int d^3 x \; x^2 \, \rho_{\rm ch} (\vx )  = \sum_{i = 1}^A \hat{p}_i\; \vx_i^{\,2} + Z\; \langle r^2 \rangle_p + N\; \langle r^2 \rangle \nr2 \, , \eqno({\rm  B5c})
$$
where $\langle r^2 \rangle_p = \int d^3 y \; y^2 \, \rho_p (y)$ and  $\langle r^2 \rangle \nr2 = \int d^3 y \; y^2 \, \rho_n (y)$ are the mean-square charge radii of the proton and neutron, respectively. The quadrupole operator is given by
$$
Q^{\alpha \beta} = \int d^3 x \; (\vx^{\,\alpha} \vx^{\,\beta} - \vx^2 \delta ^{\alpha \beta}/3 ) \, \rho_{\rm ch} (\vx ) = \sum_{i = 1}^A \hat{p}_i \; (\vx_i^{\,\alpha}\vx_i^{\,\beta}- \vx_i^2 \delta^{\alpha \beta}/3 ) \, . \eqno({\rm  B5d})
$$ 
The final charge operator that we require is the retarded dipole operator
$$\eqalignno{
\vO = \int d^3 x \; \vx\; \vx^2 \, \rho_{\rm ch} (\vx)  = &\sum_{i = 1}^A \left [\hat{p}_i\; \vx_i\, \vx_i^{2} + \frac{5}{3}  \frac{\tau_z (i)}{2} \vx_i  \left ( \langle r^2 \rangle_p - \langle r^2 \rangle \nr2 \right ) \right ] \cr
=&  \sum_{i = 1}^A \hat{p}_i\; \vx_i\, \vx_i^{2} + \frac{5}{3}  \vD \left ( \langle r^2 \rangle_p - \langle r^2 \rangle \nr2 \right )\cr
\equiv& \vO_0 + \frac{5}{3} \vD \left ( \langle r^2 \rangle_p - \langle r^2 \rangle \nr2 \right )\, , &({\rm  B5e})
}
$$
where we have used Eqn.~(B5b). Thus the mean-square radii of the proton and neutron play a role in the mean-square-radius and retarded-dipole operators, but not in the dominant (unretarded) dipole operator, or in the quadrupole operator.

The mean-square charge radius of the deuteron \cite{rsq} in zero-range approximation is obtained by combining Eqns.~(B3) and (B5c) 
$$
\langle r^2 \rangle{\raisebox{-0.6ex}{\scriptsize \rm ch}} {\!\!\!\!{\raisebox{1.0ex}{\scriptsize \rm zr}}} = \frac {A_S^2}{16 \kappa^3} + \langle r^2 \rangle_p +  \langle r^2 \rangle \nr2 \; , \eqno({\rm B6})
$$
and is accurate to within about 2\% \cite{rsq}. Note that we have used correctly normalized wave functions to evaluate  the {\it nucleon} charge-radius terms.

\subsection*{Charge Correlation Functions}

In Section (2) we demonstrated that the sum of the (third) elastic and inelastic Zemach moments is a relatively simple correlation function given by
$$
\langle 0 |\, |\vx-\vy|^3 | 0\rangle_{\rm ch} \equiv \int  d^3 x \int d^3 y\, \langle 0 | \rho_{\rm ch} (\vy)  \rho_{\rm ch} (\vx) |0 \rangle 
\; |\vx - \vy|^3 \, . \eqno({\rm B7})
$$
We will use Eqn.~(B4) to expand the product of charge operators for a general nucleus, and then restrict ourselves to the deuteron case. Because the charge operators at a point $\vx$ are functions of the distance from that point to the coordinate of nucleon $i$, we change integration variables, $\vx \rightarrow \vx + \vx_i$. This removes the coordinate $\vx_i$ from $\rho_{\rm ch} (\vx)$, and we similarly transform $\rho_{\rm ch} (\vy)$. 
Defining $\vx_{ij} = \vx_i - \vx_j$ we can then write
$$\eqalignno{
\int  d^3 x \int & d^3 y\; \rho_{\rm ch} (\vy) \rho_{\rm ch} (\vx ) \; |\vx - \vy|^3 
= \int d^3 x \int d^3 y\;  \sum_{i,j = 1}^A |\vx - \vy + \vx_{ij}|^3 \cr
&\times (\hat{p}_i \;  \rho_p (y) + \hat{n}_i \rho_n (y))(\hat{p}_j \;  \rho_p (x) + \hat{n}_j \rho_n (x))\; \, . & ({\rm B}8)
}
$$
The proton and neutron projectors $\hat{p_i}$ and $\hat{n}_i$ are true projection operators in the sense that $\hat{p}_i^2 = \hat{p}_i$ and $\hat{n}_i^2 = \hat{n}_i$, that $\hat{p}_i\hat{n}_i = 0$, and also that these projectors commute for $i \neq j$. Using these properties we split the sum into $i=j$ and $i \neq j$ parts and obtain
$$\eqalignno{
\int & d^3 x \int d^3 y\; \rho_{\rm ch} (\vy) \rho_{\rm ch} (\vx ) \; |\vx - \vy|^3 =  Z \langle r^3 \rangle_{(2)}^{p p}  + N \langle r^3 \rangle_{(2)}^{n n} \cr
+& \int  d^3 x \int d^3 y\; \sum_{i \neq j = 1}^A (\hat{p}_i \;  \rho_p(y) + \hat{n}_i \rho_n(y))(\hat{p}_j \;  \rho_p(x) + \hat{n}_j \rho_n(x))\; |\vx - \vy + \vx_{ij}|^3 \, ,
\cr
& & ({\rm B}9)
}
$$
where $\langle r^3 \rangle_{(2)}^{p p}$  and $ \langle r^3 \rangle_{(2)}^{n n}$ are the usual third Zemach moments for protons and neutrons, respectively. Although these moments vanish for point-like nucleons,  the muon's Coulomb force can interact with different parts of the charge distribution of an extended nucleon, which results in a Zemach moment.

The $i \neq j$  term above can be re-expressed in terms of familiar densities. Changing variables by $\vx \rightarrow \vx + \vy$ and performing the $\vy$ integral leads to
$$\eqalignno{
 \int  d^3 x &\int d^3 y\; \sum_{i \neq j = 1}^A (\hat{p}_i \;  \rho_p(y) + \hat{n}_i \rho_n(y))(\hat{p}_j \;  \rho_p(x) + \hat{n}_j \rho_n(x))\; |\vx - \vy + \vx_{ij}|^3 \cr
= &  \sum_{i \neq j = 1}^A  \int  d^3 x \; |\vx  + \vx_{ij}|^3 \left(\hat{p}_i \;\hat{p}_j\, \rho_{(2)}^{pp} (x)  + \hat{n}_i \;\hat{n}_j\, \rho_{(2)}^{nn} (x)  + 2 \hat{p}_i \;\hat{n}_j\, \rho_{(2)}^{pn} (x) \right) \cr
&\longrightarrow  2 \int  d^3 x \; |\vx  + \vr |^3 \rho_{(2)}^{pn} (x) \equiv 2\, C^{\,pn} (r) \, , & ({\rm B}10)
}
$$
where $\rho_{(2)}^{pn} (x) = \int d^3 y\, \rho_p(|\vx +\vy|)\, \rho_n(y)$ is the Zemach charge density \cite{annals} for {\sl overlapping} proton and neutron distributions, etc., and the arrow indicates the result for the deuteron case. In the deuteron both nucleons cannot be protons (or neutrons) and the corresponding terms vanish, although they do not vanish for He. The correlation function $C^{\,pn}(r)$ results when the muon's charge interacts with overlapping proton and neutron charge distributions whose centers are separated by a distance $r$. We finally find for the deuteron
$$
\langle 0 |\, |\vx-\vy|^3 | 0 \rangle_{\rm ch}^d  =  \langle r^3 \rangle_{(2)}^{p p}  + \langle r^3 \rangle_{(2)}^{n n} + 2\, \langle 0 | C^{\,pn} (r) | 0 \rangle \, . \eqno({\rm B}11)
$$
We will use the value $\langle r^3 \rangle_{(2)}^{p p} = 2.71(13)$ fm$^3$ \cite{pzemach}, and will ignore the very small neutron  Zemach moment. The remaining two terms in Eqn.~(B11) are rather small and largely cancel.

The proton Zemach moment is known experimentally. The remaining term in Eqn.~(B11) requires a model in order to construct $C^{\,pn} (r)$, although its leading  and  most important term for large $r$  is model independent. The required effort is substantial and tedious, but fortunately has already been performed in Appendix A of Ref.~\cite{hyper} for reasonable (but certainly improvable) models of the proton and neutron form factors. In that work we chose a dipole form factor for the proton, which has a single length parameter and generates an exponential charge distribution: $\rho_p (x) = \exp{(-\beta x)}\, \beta^3/8 \pi $. For the neutron we chose a modified Galster form factor \cite{galster} that produces a similar form with the same length parameter: $\rho_n (x) =  \lambda \beta^5 \exp{(-\beta x)}\, (3- \beta x)/32 \pi$. This density has a vanishing volume integral, generates a form factor that rises with slope $\lambda$, and therefore has a mean-square radius of $-6 \lambda$ (the conventional negative sign reflects a rising rather than falling form factor as momentum transfer increases). We use the values  $\beta$ = 4.12 fm$^{-1}$ that corresponds to $\langle r^2 \rangle_p^{1/2} = \sqrt{12}/\beta$ = 0.841 fm \cite{PSI}, and $\lambda$ = 0.01935(37) fm$^{2}$ that corresponds to $\langle r^2 \rangle_n = -$0.1161(22) fm$^2$ \cite{pdg}. We can then immediately calculate another quantity that we require: $\langle r^2 \rangle_p - \langle r^2 \rangle_n $= 0.8232(23) fm$^2$.

 Reference~\cite{hyper} calculated the quantities $\rho_{(2)}^{pn} (x)$ (called $\rho_{DG}$ in its Eqn.~(A5)) and $C^{\,pn} (r)$ (called $C^{\prime}_{DG}$ in its Eqn.~(A10)). An important feature of the latter is the behavior of the leading term for large $\beta r$: $C^{\,pn} (r) \rightarrow -12 \lambda r \,+ $vanishing terms. Equation~(D11b) of Ref.~\cite{annals} displays the behavior of $|\vx + \vr|^3$ required for constructing $C^{\,pn} (r)$ in our Eqn.~(B10). It demonstrates that for large $r$ the coefficient of $r$  is twice the mean-square radius of $\rho_{(2)}^{pn} (x)$ (viz.,  $-12 \lambda$)  and is determined entirely by the slope of the neutron form factor (viz., the measured quantity $\lambda$). It thus is a model-independent operator.

In order to be as specific as possible we rewrite Eqn.~(B11) and separate $C^{\,pn} (r)$ into the model-independent part ($- 12 \lambda r$) and a model-dependent part that  we call $g(r)$
$$
\langle 0 | |\vx-\vy|^3 |0 \rangle_{\rm ch}^d  =  \left ( \langle r^3 \rangle_{(2)}^{p p}   -24\, \lambda\, \langle 0 | r | 0 \rangle \right ) + 2\, \langle 0 |  g (r) | 0 \rangle \, , \eqno({\rm B}12)
$$
where the  two terms in parentheses are dominant and nucleon-model independent. Equation~(B3) can be used to estimate the second term
$$
-24\, \lambda\, \langle 0 | r | 0 \rangle\zrlow = - \frac{6\, \lambda\, A_S^2}{\kappa^2} \, , \eqno({\rm B}13)
$$
while the model-dependent term can be similarly estimated
$$
2\, \langle 0 |  g (r) | 0 \rangle\zrlow \cong \frac{24 \lambda A_S^2}{\beta^2} \left(10 \ln{(2 \kappa/\beta)} + \frac{77}{12} + \cdots \right) \, , \eqno({\rm B}14)
$$
and depends only weakly on $\kappa$. The model parameter $\beta$ is roughly 20 times $\kappa$, which explains the dominance of the model-independent term over the model-dependent one.

\subsection*{Energy-Weighted Sum Rules}

The remaining operators are transition operators that connect the deuteron ground state to either plane-wave excited states or phase-shifted free waves that are parameterized by asymptotic scattering properties such as s-wave scattering lengths (denoted $a$ below). For s-waves the parameter $\kappa a$ determines the importance of the asymptotic modification.
Various energy weightings of the squared matrix elements are then summed. In the simplest form the excited states are $| N \rangle = | \vk \rangle = e^{i \vk \cdot \vr}$, $\omega_N = (\vk^2 + \kappa^2)/2 \mu$ is the difference in energy between the $N\underline{th}$ final state, $E_N = \vk^2/2 \mu$, and the ground state, $E_0 = -\kappa^2/2 \mu \equiv - E_B$. The corresponding phase space is $d^3 k/(2 \pi)^3$ (i.e., $\sum_N = \int d^3 k/(2 \pi)^3$). The quantity $\mu$ is the n-p reduced mass.

\subsubsection*{Dipole Sum Rules}

None of the operators in Eqns.~({\rm B5}) involve spin, and transitions lead only to spin-triplet states. We thus only require isospin matrix elements between the isoscalar deuteron (s-wave) and isovector negative-parity excited states (viz., p-wave) or isoscalar positive-parity excited states (viz., s-wave or d-wave). The isospin matrix element of the large-bracketed isovector operator in Eqn.~(B5b) between the isospin-0 ground state and any isospin-1 excited state is 1, leaving only the factor of $\vr/2$ to treat. This produces
$$
\langle N | \vD | 0 \rangle\zrlow = \langle \vk |\, \vr/2\,  | 0  \rangle\zrlow = \int_0^{\infty} d^3 r\; e^{-i \vk \cdot \vr}\, \left[\frac{\vr}{2}\right] \left[ \frac{e^{-\kappa r}}{r} \frac{A_S}{\sqrt{4 \pi}} \right] =  \frac{-i \vk \sqrt{4 \pi} A_S}{(k^2 + \kappa^2)^2}\, . \eqno({\rm B15})
$$
Using this result any energy-weighted dipole sum rule can be constructed in zero-range approximation provided that $p > -3/2$
$$\eqalignno{
S_p^D =& \sum_N \frac{|\langle N | \vD | 0 \rangle\zrlow |^2}{\omega_N^p} = \frac{4 \pi A_S^2}{(2 \pi)^3} \int d^3 k \frac{k^2 (2 \mu)^p}{(k^2 + \kappa^2)^{4+p}} \cr
=&  \left [ \frac{3\, \Gamma(3/2+p)}{4 \sqrt{\pi}\, \Gamma (4+p)} \right ] \left [\frac{A_S^2}{\kappa^{3}E_B^p} \right ]
\longrightarrow  \left [ \frac{A_S^2}{2 \pi \kappa^{3}E_B^{1/2}} \right ] \left[ \frac{8}{35} \right]\, , &({\rm B16a})
}
$$
where $E_B = \kappa^2/2 \mu \cong $ 2.2 MeV sets the energy scale for these sum rules. The arrow points to the $p = 1/2$ case that determines the leading-order  polarization correction. We also require the $p=-1/2$ case
$$
S_{-1/2}^D =   \left[ \frac{A_S^2 E_B^{1/2}}{2 \pi \kappa^{3}}\right] \left[ \frac{4}{5} \right] \, . \eqno({\rm B16b})
$$
The vanishing of p-waves at the origin necessarily enhances the quality of dipole sum rules in  zero-range approximation.

The deuteron electric polarizability is given by $\alpha{\raisebox{-0.6ex}{\tiny \rm E}} = 2 \alpha S^D_1 / 3 = \alpha \mu A_S^2/32 \kappa^5$, and is approximately 3/4\% too large compared to potential models \cite{elpol,dipole}. Corrections from p-wave scattering volumes are ${\cal{O}}(1/\kappa^2)$, while d-wave corrections and short-range s-wave corrections are both ${\cal{O}}(1/\kappa)$ \cite{elpol}.

The overlap of the unretarded and retarded dipole matrix elements is the most important correction to the usual dipole sum rules \cite{myhe4}. From Eqn.~(B5e) we calculate
$$\eqalignno{
\langle N | \vO_0 | 0 \rangle\zrlow =& \langle \vk |\, \vr\, r^2/8\,  | 0  \rangle\zrlow = \int_0^{\infty} d^3 r\; e^{-i \vk \cdot \vr}\, \left[\frac{\vr\, r^2}{8}\right] \left[ \frac{e^{-\kappa r}}{r} \frac{A_S}{\sqrt{4 \pi}} \right]\cr
=&  \frac{-i \vk \sqrt{4 \pi} A_S\, (5 \kappa^2 - k^2)}{(k^2 + \kappa^2)^4}\, . &({\rm B17})
}$$ 
Combining terms we form the retarded E1 sum rule for $p > -5/2$
$$\eqalignno{
\Delta S_p^{\rm E1} =& \sum_N \frac{\langle N | \vD | 0 \rangle\sstar\!\!\zrlow \cdot \langle N | \vO_0 | 0 \rangle \zrlow}{\omega_N^p} 
= \frac{4 \pi A_S^2 (2 \mu)^p}{(2 \pi)^3} \int d^3 k\; \frac{k^2(5 \kappa^2 - k^2)}{(k^2 + \kappa^2)^{6+p}} \cr
=&  \left [ \frac{15\; \Gamma(5/2+p)(2+p)}{4 \sqrt{\pi}\, \Gamma (6+p)} \right ] \left [\frac{A_S^2}{\kappa^{5}E_B^p} \right ] {\longrightarrow}\left [\frac{A_S^2\, E_B^{1/2}}{2 \pi \, \kappa^{5}} \right ] \left[\frac{8}{21}\right]  \, , &({\rm B18})
}$$
where the last result holds for $p = -1/2$, which we require. We can combine Eqns.~(B5e), (B16) and (B18) to yield the sum rules
$$
\sum_N \frac{\langle N | \vD | 0 \rangle\sstar\!\!\zrlow \cdot \langle N | \vO | 0 \rangle \zrlow}{\omega_N^p} = \Delta S_p^{\rm E1} + \frac{5}{3} S_p^{\rm D} \left ( \langle r^2 \rangle_p - \langle r^2 \rangle \nr2 \right )\, ,\eqno ({\rm B19})
$$
which contain a contribution from finite nucleon size. We require $p = -1/2$.

\subsubsection*{Quadrupole Sum Rules}

Quadrupole excitations are generated by
$$\eqalignno{
\langle N | Q^{\alpha \beta}| 0 \rangle\zrlow =& \langle \vk |\, (r^{\alpha}r^{\beta} - r^2 \delta^{\alpha \beta}/3)/4\,  | 0  \rangle\zrlow\cr
 =& \int_0^{\infty} d^3 r\; e^{-i \vk \cdot \vr}\, \left[\frac{r^{\alpha}r^{\beta} - r^2 \delta^{\alpha \beta}/3}{4}\right] \left[ \frac{e^{-\kappa r}}{r} \frac{A_S}{\sqrt{4 \pi}} \right]\cr
=&  \frac{-2  \sqrt{4 \pi} A_S (k^{\alpha}k^{\beta} - k^2 \delta^{\alpha \beta}/3)}{(k^2 + \kappa^2)^3}\, . &({\rm B20})
}$$ 
The corresponding quadrupole sum rules are given for $p > -5/2$ by
$$\eqalignno{
S_p^Q =& \sum_N \frac{|\langle N |Q^{\alpha \beta}  | 0 \rangle\zrlow |^2}{\omega_N^p} = \frac{8}{3}\cdot\frac{4 \pi A_S^2}{(2 \pi)^3} \int d^3 k \frac{k^4 (2 \mu)^p}{(k^2 + \kappa^2)^{6+p}}\cr
=&  \left [ \frac{5\, \Gamma(5/2+p)}{\sqrt{\pi}\, \Gamma (6+p)} \right ] \left [\frac{A_S^2}{\kappa^{5}E_B^p} \right ] {\longrightarrow}\left [\frac{A_S^2\, E_B^{1/2}}{2 \pi \, \kappa^{5}} \right ] \left[\frac{64}{189}\right] \, , &({\rm B21})
}$$
where the last result holds for $p= -1/2$, which we require.

\subsection*{Sum Rules Involving S-Wave Transitions}

Because total charge is a super-conserved quantity it does not generate transitions, and therefore only the first term in the mean-square-radius operator in Eqn.~(B5c) generates the leading monopole excitations. In the deuteron case this means $^3S_1 \rightarrow  {^3}S_1$ transitions. Magnetic interactions in nuclei are dominated by spin-flip transitions, and in the deuteron case this means $^3S_1 \rightarrow  {^1}S_0$  transitions. At very low energies the scattering length determines the form of both the scattered wave function and the scattering amplitude. Both the triplet scattering length, $a_t =  $5.4194(20) fm \cite{deut}, and the singlet scattering length, $a_s = -23.748(10)$ fm \cite{dumbrajs}, are very large and could significantly impact zero-range calculations. The technique for treating these cases was developed in Refs.~\cite{khrip,yulik} and we closely follow that treatment.

The ${^1\!}S_0$ state is a ``virtual'' state, characterized by a pole on the imaginary axis in the lower half of the analytic k-plane at $k_v = \frac{i}{a_s}$, where the singlet scattering length $a_s$ is large and negative. This pole is very close to the origin at an energy $E_v = \frac{k_v^2}{M_N} =-74$ keV on the second sheet of the complex energy plane. The ${^3}S_1$ state on the other hand is characterized by the deuteron bound-state pole at $k = i\, \kappa$, which greatly affects that scattering length.  Reference \cite{yulik} astutely observes that orthogonality of the zero-range bound and zero-energy $^3S_1$ scattering  wave functions requires that $a_t = 1/\kappa$ = 4.3 fm, which is comparable to the experimental value but roughly 20\% too low.

The asymptotic form of the wave function for both s-wave excited states (denoted generically by $S\sstar$) is given by basic principles as \cite{khrip,yulik}
$$
R_{S{\raisebox{0.5ex}{\tiny *}}} (r) = \frac{\sin(k r)}{k r} + e^{i \delta}\sin{\delta}\; \frac{e^{ikr}}{kr} \longrightarrow \frac{\sin(k r)}{k r} - \frac{a}{1+ika} \frac{e^{ikr}}{r}\, , \eqno({\rm B22})
$$
where we have used $k \cot{\delta} \cong -1/a$ to arrive at the final form. This is an approximation that ignores the effective-range and higher corrections and is valid only at very low energies. 

\subsubsection*{Monopole Sum Rules}

The $^3S_1 \rightarrow  {^3}S_1$ monopole excitations can now be calculated using the first term in the mean-square-radius operator in Eqn.~(B5c):
$$\eqalignno{
\langle N | \hat{r}^2| 0 \rangle\zrlow =& \langle {^3}S_1 |\,  r^2 /4\,  | 0  \rangle\zrlow
 = \int_0^{\infty} d^3 r\; R\sstar_{\!\!{^3\!}S_1} (r)\, \left[\frac{r^2}{4}\right] \left[ \frac{e^{-\kappa r}}{r} \frac{A_S}{\sqrt{4 \pi}} \right]\cr
=&  \frac{\sqrt{\pi} A_S \left(3 \kappa^2 - k^2 +(\kappa a_t) (3 k^2 - \kappa^2) \right)}{(k^2 + \kappa^2)^3 (1-i k a_t)} \cr
=&  \frac{\sqrt{\pi} A_S \left(\bar{a}\,(\kappa^2 + k^2) +\bar{b} \kappa^2\right)}{(k^2 + \kappa^2)^3 (1-i k a_t)}\,, &({\rm B23})
}$$ 
where $\bar{a} = 3\kappa a_t -1$ and $\bar{b} = 4(1 - \kappa a_t)$.
This leads to the energy-weighted sum rules for $p > -7/2$
$$\eqalignno{
S_p^{\,\hat{r}^2} =& \sum_N \frac{|\langle N | \hat{r}^2 | 0 \rangle\zrlow |^2}{\omega_N^p}\cr =& \frac{\pi A_S^2  (2 \mu)^p}{(2 \pi)^3} \int d^3 k\; \frac{\bar{a}^2(k^2 + \kappa^2)^2 +2 \bar{a}\bar{b} \kappa^2 (k^2 + \kappa^2) +\bar{b}^2 \kappa^4}{(k^2 + \kappa^2)^{6+p}\; (1+k^2 a^2)}\cr
=&  \left [\frac{A_S^2}{2 \pi \kappa^{5} E_B^p} \right ]\,\left[ \bar{a}^2 I_{4+p} +2 \bar{a} \bar{b}\, I_{5+p} +\bar{b}^2\, I_{6+p} \right], &({\rm B24})
}$$
where we have used identity {\sl 3.197.5} of Ref.~\cite{GR} to define
$$\eqalignno{
I_{\alpha} (y) =& \int_0^{\infty} dx\; \frac{x^2}{(1+x^2)^{\alpha}\; (1+x^2 y^2)}\cr
 =& \frac{\sqrt{\pi} \Gamma(\alpha -1/2)}{4 \Gamma(\alpha+1)}  {_2}F_1 (1,3/2,\alpha+1;1-y^2)  & ({\rm B25})
}
$$
in terms of $y = \kappa a_t$ and a Gauss hypergeometric function.

As a practical matter we require only $p = - 1/2$, or $\alpha = 7/2, 9/2, 11/2$. Writing $\alpha = 1/2 +m$, we therefore require $m= 3, 4, 5$  and we shall see below that we also need $m=1$ for magnetic sum rules. This leads to
$$\eqalignno{
I_{1/2 + m}\, (y) =& \frac{2^{m-1}\, (m-1)!}{(2 m + 1)!!}\;  {_2}F_1 (1,3/2,3/2 + m;1-y^2) \cr
\equiv& G_m (1-y^2) \, , &({\rm B26})
}
$$
where $G_m (z)$ is defined in Eqn.~(C10). In Appendix C we develop useful representations of ${_2}F_1 (1, 3/2, 3/2 + m; z)$ for $z \leq 1$ and integer $m \geq 1$   in terms of logarithms and powers of $z$ (viz., $G_m (z)$). This leads to a simplified form for the particular variant of Eqn.~(B24) that we require:
$$
S_{- 1/2}^{\,\hat{r}^2} =  \left [\frac{A_S^2 \, E_B^{1/2}}{2 \pi \kappa^{5}} \right ]\,\left[ \bar{a}^2 G_3 +2 \bar{a} \bar{b}\, G_4 +\bar{b}^2\, G_5 \right], \eqno({\rm B27})
$$
Equations~(C6), (C8), and (C10) can be used to evaluate the three $G_m$. Equation~(C13) can be used to show that the small square bracket in Eqn.~(B27)  equals 106/315 for vanishing $a_t$ and $y$ (with $\bar{a} \rightarrow -1$ and $\bar{b} \rightarrow 4$).

\subsubsection*{Magnetic Sum Rules}

Deuteron magnetic properties can be treated analogously, and we again closely follow the treatment in Refs.~\cite{khrip,yulik}. The nuclear magnetic-moment operator is given by
$$
\vmu = \sum_{i = 1}^A \left( \frac{\hat{\mu} (i) \, \vsig (i) + 
\hat{p}_i \, {\vL} (i)}{2 M} \right) + \vmu_{MEC} \, , \eqno ({\rm B28})
$$
where $\vsig (i)$ is the (Pauli) spin operator of nucleon $i$, $ {\vL} (i)$ is the orbital angular momentum of nucleon $i$, $\vmu_{MEC}$ is the contribution of meson-exchange (primarily pion-exchange) currents, and the spin-magnetization current of nucleon $i$ is determined by
$$
\hat{\mu} (i) = \mu_p \, \hat{p}_i + \mu_n \, \hat{n}_i \, . \eqno ({\rm B29})
$$
The isoscalar and isovector combinations of the proton and neutron magnetic moments are very
different in size: $\mu_s \equiv \mu_p + \mu_n = 0.8798\cdots$ and 
$\mu_v \equiv \mu_p - \mu_n = 4.7059\cdots$.  The large isovector magnetic moment (corresponding to a $^3S_1 \rightarrow  {^1}S_0$ transition) completely dominates, and we will ignore for now the isoscalar combination (corresponding to a $^3S_1 \rightarrow  {^3}S_1$ transition). We note that the orbital contribution vanishes for an s-wave deuteron, and we also ignore the  meson-exchange contribution because it has the same range as nuclear potentials (although it enhances by roughly 15\% \cite{higher,MEC}).

The spin and isospin matrix elements are easily performed for a $^3S_1 \rightarrow  {^1}S_0$ transition and we find
$$
\langle {^1}S_0 | \mu_z | 0 \rangle\zrlow = \frac{\mu_v}{2\, M_N} \int d^3 r\; R\sstar_{\!\!{^1\!}S_0} (r)\, \left[ \frac{e^{-\kappa r}}{r} \frac{A_S}{\sqrt{4 \pi}} \right]\, . 
\eqno({\rm B30})
$$
Using Eqn.~(B22) for $R\sstar_{\!\!{^1\!}S_0} (r)$ and performing the integral we obtain
$$
| \langle {^1}S_0 |\, \vmu \,| 0 \rangle\zrlow |^2 = \frac{\pi A_S^2\, \mu_v^2\, (1-\kappa a_s)^2}{ M_N^2\, (k^2+\kappa^2)^2 \,(1+k^2 a_s^2)} \, ,
\eqno({\rm B31})
$$
and note that the very large scattering length causes the matrix element to decrease very rapidly \cite{dphoto} for $E = \frac{k^2}{M_N} \rsim |E_v|$.

Magnetic sum rules analogous to Eqn.~({\rm B16}) can be defined for $p > - 3/2$ 
$$\eqalignno{
S_p^{\mu} =& \sum_N \frac{|\langle {^1}S_0 | \vmu | 0 \rangle\zrlow |^2}{\omega_N^p}\cr
 =& \frac{\pi A_S^2\, \mu_v^2\, (1-\kappa a_s)^2}{M_N^2\, (2 \pi)^3} \int d^3 k \frac{(2 \mu)^p}{(k^2 + \kappa^2)^{2+p}\, (1+k^2 a_s^2)}\cr =& \frac{A_S^2\, \mu_v^2\, (1-\kappa a_s)^2 (2 \mu)^p}{\, 2 \pi\, \kappa^{1+2p} M_N^2} \int_0^{\infty} dx \frac{x^2}{(1+x^2)^{2+p}\,(1+x^2 y^2)}  \cr
=&   \frac{A_S^2\, \mu_v^2\, (1-y)^2}{2 \pi \kappa\, E_B^p M_N^2} I_{2+p} (y)  , &({\rm B32})
}$$
where $y = \kappa a_s$ and we have used Eqn.~(B25). Convergence for $y=0$ requires $p > -1/2$.

The deuteron magnetic susceptibility  corresponds to $\beta_d = 2 \alpha S_1^{\mu}/3$ (i.e., $p=1$), and we find using $2\mu = M_N$
$$
\beta_d = \frac{\alpha\, A_S^2\, \mu_v^2\, (|y| + 1/3) (y-1)^2}{16\, \kappa^3 M_N (1+|y|)^3}\, , \eqno({\rm B33})
$$
where the integral $I_3$ is straightforward to evaluate using partial fractions. This agrees with Ref.~\cite{khrip} if we use $A_S^2 \rightarrow 2 \kappa$ (i.e., $|\psi_S|^2$ is normalized to 1).

Our magnetic polarizability sum rule corresponds to $p=-1/2$ or $m=1$ in the notation of Eqn.~(B26)
$$
S_{-1/2}^{\mu} = \frac{A_S^2\, \mu_v^2}{2 \pi \, M_N^{5/2}} \, \left[(1-y)^2 G_1 (1-y^2) \right] \,  ,\eqno({\rm B34})
$$
and is only interesting to us because of the very large value of $\mu_v^2$.

The isoscalar contribution to the magnetic polarizability sum rule in zero-range approximation is obtained formally by replacing $\mu_v^2$ by the much smaller $2 \mu_s^2$ in Eqn.~(B32), and replacing $a_s$ in $y$ by the spin-triplet scattering length $a_t = $5.4194(20) fm \cite{deut}. This should vanish for the contributions from the spin and orbital angular momentum to the magnetic moment operator because exact radial wave functions of the ground state and excited states are orthogonal. Unfortunately that orthogonality of the radial wave functions in zero-range approximation obtains {\sl only} for $y = 1$ or $a_t = 1/\kappa$ \cite{yulik}, and is therefore only approximate for the physical value of $a_t$, corresponding to y=1.2552. The result is nevertheless greatly suppressed and  will be ignored.

\subsection*{Gauge Sum Rules}

We collect here six higher-order terms that were not treated above, two each from the charge, current, and seagull parts.  We first examine terms that involve sum rules with powers of $\omega_N/m$ less than 3/2. This criterion omits terms that are suppressed by at least a factor of $(\omega_N/m)^2 \sim 1/400$, {\sl provided that the sum rules saturate at low excitation energies}. With increasing powers of $\omega_N$,  matrix elements become more and more sensitive to the deuteron's short-range behavior, and also saturate at higher and higher energies.  They eventually diverge in zero-range approximation because the deuteron wave function diverges at short range. Logarithmically divergent terms can still be roughly estimated using a cutoff.

Sum rules linear in $\omega_N$ are crucial for maintaining the gauge invariance of the underlying nuclear Compton amplitude in Fig.~(1). This is discussed in some detail in Appendix B of Ref.~\cite{higher}. They can be evaluated using a trick involving commutators and closure. For $n \geq 2$ we define a sum rule
$$\eqalignno{
S_n^C &= \sum_{N \neq 0}  \int  d^3 x \int d^3 y\, \langle 0 | \rho_{\rm ch} (\vy) |N \rangle \langle N | \rho_{\rm ch} (\vx) |0 \rangle \; \omega_N\; |\vx - \vy|^n \cr
&=  \haf \int  d^3 x \int d^3 y\, \langle 0 |\, [ \rho_{\rm ch} (\vy) , [H, \rho_{\rm ch} (\vx)]]\, |0 \rangle \;  |\vx - \vy|^n &({\rm B35})
}
$$
that is equivalent to a single matrix element of a double commutator with the Hamiltonian, $H$.  The zero-range approximation ignores the potential part of H and only the kinetic energy part contributes. In reality the potential terms can enhance the $n=2$ sum rule by a factor approaching two \cite{MEC}, although the enhancement is less for the weakly bound deuteron \cite{Ji}. Performing the commutators for point-like nucleons without the potential produces
$$
S_n^C = \frac{n(n+1)}{4 M_N} \int  d^3 x \int d^3 y\, |\vx - \vy|^{\,n-2} \langle 0 | \rho_{\rm ch} (\vy)  \rho_{\rm ch} (\vx) |0 \rangle =0 \eqno({\rm B36})
$$
 in the deuteron for $n > 2$ and $S^C_2 = -3/2 M_N$ for $n=2$. The vanishing point-nucleon result obtains because the charge must reside on the single proton and thus $\vx - \vy = 0$. This explains the substantial cancellations that occur in the final term in Eqn.~(6) when expanded in partial waves (even though the sum rule involves a factor of $\sqrt{\omega_N}$ rather than $\omega_N$). Equation~(B36) also causes the final term in Eqn.~(A16) to vanish.

The constant (i.e., $z$-independent) terms in $J_N$ and $K$ and the $z^2 \omega_N$ term in Eqn.~(A16) combine to produce a linear-in-$\omega_N$ term that is quite small ($\sim -0.001$ meV) in zero-range approximation because of cancellations between the terms. Note that we use only the inelastic part of the seagull in order to cancel the infrared divergences, as discussed in Ref.~\cite{higher}. The terms in $J_N$ and $\bar{J}_N$ that are quadratic in $\vz$ (except for the M1 term) are logarithmically divergent, but can be estimated using a cutoff. They scale like the prefactor in Eqn.~(12) times $A_S^2/\pi \sqrt{2 m} M_N^{5/2}$ times a number on the order of one.  Taking that number to be one, we find a contribution slightly less than 0.001 meV. The final contribution is from the terms in $K$ and $\bar{K}$ that are quadratic in $\vz$. If we replace $\beta$ in the logarithm by an average value, $\bar{\beta}$, we find a result that scales like the prefactor times $A_S^2\, m /128 \pi M_N\, \kappa^3$ times a number on the order of one. The result is -0.002 meV. All of these contributions are small, and have no effect on the results in Table 1. 

\subsection*{Relativistic Corrections in Deuteron}

Given the suppression of sub-dominant terms in Table 1, it should be sufficient to treat relativistic corrections only in the dominant unretarded-dipole term. This contribution is determined by the sum rule in Eqn.~(B16a), which requires only the dipole operator and the energy difference of excited and ground states.

Relativistic corrections to the deuteron dipole operator are thoroughly treated in Ref.~\cite{dphoto}. The electromagnetic spin-orbit interaction that generates fine-structure splitting in atoms is the most obvious source, but its dipole operator is spin dependent and cannot interfere with the usual spin-independent dipole operator. Potential-dependent dipole operators are outside the domain of the zero-range approximation. This leaves only the usual dipole operator to treat in this work. We note that one tiny correction not incorporated into our treatment is quite trivial and indeed is classical. The dipole moment is the distance from the deuteron's CM to the center of the proton. Because the neutron's mass, $m_n$,  is slightly greater than the proton's mass, $m_p$  (by $\sim$ 1.3 MeV), the dipole operator is given by $\frac{m_n}{(m_n+m_p)} \vr$ and is very slightly larger ($\sim 0.1\%$) than the $\vr/2$ that we use \cite{ingo}.

Given the dipole operator our relativistic dipole sum rule requires only the appropriate expression for energy differences. In the absence of any potential the CM Hamiltonian for two equal-mass nucleons is given by $2 \sqrt{\vp^2 + M_N^2}$. Plane waves are eigenfunctions of this Hamiltonian, as are bound-state wave functions of the generic form: $\exp{(-\kappa_r r)}/r$ (for $r \neq 0$). Although these wave functions are also eigenfunctions for the non-interacting non-relativistic Hamiltonian, in the latter case the binding parameter $\kappa$ should be labeled $\kappa_{nr}$. If $E_B > 0$ is the experimentally determined deuteron binding energy, then relativistic kinematics requires that
$$
E_B = 2 M_N - 2 \sqrt{-\kappa_r^2 + M_N^2} \, , \eqno({\rm B37a})
$$
rather than the non-relativistic version 
$$
E_B = \kappa_{nr}^2/M_N \, . \eqno({\rm B37b})
$$
The two parameters satisfy the relationship
$$
\kappa_r^2 = \kappa_{nr}^2  \left(1- \frac{\kappa_{nr}^2}{4\, M_N^2} \right) \, . \eqno({\rm B38})
$$
Since relativistic corrections in the zero-range approximation are expected on dimensional grounds to be multiples of $\kappa_{nr}^2/M_N^2$, it is clearly necessary to distinguish between $\kappa_r$ and $\kappa_{nr}$ in the results. Although we haven't labeled the $\kappa$ used in this work, we note that the AV18 potential model uses Eqn.~(B37b) to determine $\kappa$ \cite{AV18,wiringa} and thus we have chosen to use $\kappa_{nr}$ in order to make detailed comparisons with Ref.~\cite{KP}.

We can now easily construct the relativistic version of $S^D_p$ in Eqn.~(B16a) by using
$$
\omega_N = 2 \sqrt{\vk^2 + M_N^2} - 2 \sqrt{-\kappa_r^2 + M_N^2} \, , \eqno({\rm B39})
$$
with $\kappa_r$ in the dipole matrix elements and using the usual phase space integral for summing over the excited states. Note that we can convert $\omega_N$ into non-relativistic {\sl form} by multiplying it by $\sqrt{\vk^2 + M_N^2} + \sqrt{-\kappa_r^2 + M_N^2}$, which approximately equals $2 M_N + (k^2 - \kappa_r^2)/2 M_N$. Expanding to leading order in $1/M_N^2$ we find for $p > -1/2$
$$\eqalignno{
S_p^{D_{\rm rel}} =& \sum_N \frac{|\langle N | \vD | 0 \rangle\zrlow |^2}{\omega_N^p} \cong \frac{4 \pi A_S^2\, (2 M_N)\hip}{(2 \pi)^3 \,  2\hip} \int d^3 k \, \frac{k^2 (1 + p(k^2 - \kappa_r^2)/4 M_N^2)}{(k^2 + \kappa_r^2)^{4+p}} \cr
=&  \left [ \frac{3\, \Gamma(3/2+p)}{4 \sqrt{\pi}\, \Gamma (4+p)} \right ] \left [\frac{A_S^2\, M_N^p}{\kappa_r^{3+2 p}} \right ]
\left(1 + \frac{p (2-p)}{2 (1+2 p)} \frac{\kappa_r^2}{M_N^2} \right) \cr
\longrightarrow & \left [ \frac{A_S^2 M_N^{1/2}}{2 \pi \kappa_r^{4}} \right ] \left[ \frac{8}{35} \right] \left(1 + \frac{3}{16} \frac{\kappa_r^2}{M_N^2} \right) \, , &({\rm B40})
}
$$
where the arrow points to the $p = 1/2$ result that we require. The relativistic correction is a tiny factor of 0.0004 or an additional and negligible 0.001 meV. Note that the deuteron's very weak binding is responsible for this tiny size. Intranuclear momenta on the scale of the pion mass ($m_{\pi} \sim 3 \kappa$) that are common in heavier nuclei would generate corrections an order of magnitude larger.

We can also determine the correction to the previously calculated \cite{dipole,chi-pt} deuteron electric polarizability, which is proportional to the $p=1$ version of the sum rule above. This scales like $A_S^2(1+ \kappa_r^2/6 M_N^2)/\kappa_r^5$. Reference \cite{chi-pt} expressed all of their results in terms of $\kappa_{nr}$, and  used $A_S^2 = 2 \kappa_{nr}/(1-\kappa_{nr} \rho_d)$. Ignoring the factor containing $\rho_d$ and using Eqn.~(B38) to convert $\kappa_{nr}$  to $\kappa_r$ changes their correction of (2/3)[$\kappa_{nr}^2/M_N^2$] to (1/6)[$\kappa_{r}^2/M_N^2$], which agrees with our result above. This does not agree, however, with the result in Ref.~\cite{dipole} (except for the scales involved). In that work we computed the relativistic form of the Green's function, and found a singular term, while stating that ``in keeping with the zero-range approximation we ignore this term.'' The missing singular term can be computed and unfortunately accounts exactly for the difference between Refs.~\cite{dipole} and \cite{chi-pt} (and the present work). Dropping the singular term may have been ``in keeping with the zero-range approximation,'' but it led to an incorrect result. The result above and in Ref.~\cite{chi-pt} is correct.

We noted above that the AV18 potential model was tuned to $\kappa_{nr}$, in common with most potentials. Two versions of the Nijmegen potential models, labelled ``rel'', are tuned to Eqn.~(B37a) and thus implicitly use $\kappa_r$ in the deuteron. The electric polarizabilities for these models are indeed higher than the corresponding non-relativistic versions by the appropriate amounts, as listed in Table~(1) of Ref.~\cite{logsum}.

\section*{Appendix C - Hypergeometric Functions}

The Gauss hypergeometric function ${_2}F_1 (1, 3/2, 3/2 + m; z)$ for $z \leq 1$ and integer $m \geq 1$ can be determined from the function ${_2}F_1 (1, 3/2, 3/2; z)$ by using the identities {\sl 15.1.8} and {\sl 15.2.4} of Ref.~\cite{AS}:
$$
{_2}F_1 (1, 3/2, 3/2; z) = \frac{1}{1-z} \, , \eqno({\rm C1})
$$
where for now we assume that $z$ is positive, and 
$$
\frac{\!\!d^{m}}{d z^m} \left[z^{1/2+m}\right] {_2}F_1 (1, 3/2, 3/2 + m; z) = \frac{(2 m +1)!!\;  z^{1/2}}{2^m (1-z)} \, . \eqno({\rm C2})
$$
Equation~(C2) can be integrated $m$ times on $[0,z<1]$  to produce with the aid of identity {\sl 4.631} of Ref.~\cite{GR}
$$\eqalignno{
{_2}F_1 (1, 3/2, 3/2 + m; z) &= \frac{ (2 m + 1)!! }{(m-1)!\,2^m \, z^{1/2+m}} \int_0^z dx \, \frac{(z-x)^{m-1}\, x^{1/2}}{(1-x)} \cr
&=\frac{ (2 m + 1)!! }{(m-1)!\,2^m } \int_0^1 dx \, \frac{(1-x)^{m-1}\, x^{1/2}}{(1- z\, x)}\, , &({\rm C3})
}
$$
where the latter is a standard integral representation of this ${_2}F_1$.
The apparent singularity in the integrand at $x=1$ can be removed for $m > 1$ by 
subtracting and adding $(z-1)^{m-1}$ from $(z-x)^{m-1}$ in the numerator of the integrand:
$$
 \int_0^z dx \, \left[ \left( (z-x)^{m-1} - (z-1)^{m-1}\right) + (z-1)^{m-1}\right]\, \frac{x^{1/2}}{(1-x)} \, . \eqno({\rm C4})
$$
The last term in the square brackets multiplies an elementary integral (let $x = y^2$), which is also the complete result for $m=1$
$$
\int_0^z dx \, \frac{ x^{1/2}}{(1-x)} = 2 \sqrt{z}\; (L(z)-1) \, , \eqno({\rm C5})
$$
where the function $L(z)$ is given for both positive and negative $z$ by
$$
L(z) = \left\{ \begin{array}{lll}
                    \frac{1}{2 \sqrt{z}} \ln{\left| \frac{1+\sqrt{z}}{1-\sqrt{z}}\right|} \quad \quad& 0 \le z \le 1\\
                    \\
                    \frac{1}{\sqrt{-z}} \tan^{-1} {\scriptstyle (\sqrt{-z})} & z \le 0
                    \end{array}
           \right.
         \eqno({\rm C6})
$$
We have made the obvious extension of $L(z)$ for negative $z$ in accordance with the power series of the original ${_2}F_1$ function and those in Eqn.~(C6).

Expanding the remaining quantity in the numerator of Eqn.~(C4) as a series in $x$ and performing the integral term-by-term leads to
$$
{_2}F_1 (1, 3/2, 3/2 + m; z) = \frac{ (2 m + 1)!! }{2^{m-1} \, z^{m}}\left[ \frac{(z-1)^{m-1}}{\!\!\!\!\!(m-1)!}\,  (L(z)-1) + P_m (z)\right] \, , \eqno({\rm C7})
$$
where $P_m (z)$ is a finite series in $z$ of length $m-1$ given by
$$\eqalignno{
\!\!P_m (z) =&  \sum_{k=0}^{m-2}\, \frac{(-1)^k}{(k+1)! (m-2-k)!} \;\sum_{l=0}^{k}\, \frac{z^{m-1-l}}{2k - 2l +3} \cr
 =& \sum_{l=0}^{m-2}\, (-1)^l z^{m-1-l} \sum_{j=0}^{m-2-l} \frac{(-1)^j}{(j+l+1)! (m-2-j-l)!  (2j +3)}  . &({\rm C8})
}
$$
For ease of use in Appendix A we rewrite Eqn.~(C7) in the form
$$
{_2}F_1 (1, 3/2, 3/2 + m; z) = \frac{ (2 m + 1)!! }{2^{m-1} \, (m-1)!}\; G_m (z)  \, , \eqno({\rm C9})
$$
where
$$
G_m (z) = \left[ \frac{(z-1)^{m-1}}{\!\!\!\!\! z^{m}}\,  (L(z)-1) + \frac{(m-1)!}{z^m} P_m (z)\right] \, . \eqno({\rm C10})
$$
This result (together with Eqns.~(B25) and (B26)) was verified numerically.

For completeness we note that for $m > 1$ we have
$$
{_2}F_1 (1, 3/2, 3/2 + m; 1) = \frac{2 m + 1}{2 (m-1)} \, , \eqno({\rm C11})
$$
or equivalently
$$
P_m (1) = \frac{2^{m-2}}{(2 m -1)!! \,(m-1)} \,  , \eqno({\rm C12})
$$
and thus
$$
G_m (1) = \frac{2^{m-2}\, (m-2)!}{(2 m -1)!!} \,  . \eqno({\rm C13})
$$
Values of the energy-weighted sum rule (B27) for vanishing $a_t$ require $G_m (1) = (2/15, 8/105, 16/315)$ for $m= (3,4,5)$, respectively.

Evaluating Eqn.~({\rm C8}) for $m=1-5$ leads to
$$\eqalignno{
 P_1 (z) &= \quad 0 \cr
 P_2 (z) &=  \quad \! \frac{z}{3} \cr
 P_3 (z) &=  -\frac{z}{6} \; +\frac{7 z^2}{30} \cr
 P_4 (z) &=  \;\; \frac{z}{18} -\frac{2 z^2}{15} \;\; +\frac{19 z^3}{210} \cr
 P_5 (z) &= \! -\frac{z}{72} +\frac{17 z^2}{360} -\frac{47 z^3}{840} +\frac{187 z^4}{7560}\, , &({\rm C14})
}
$$
all of which satisfy Eqn.~(C12) for $m > 1$.

Our large s-wave scattering lengths generate fairly large values of $|y|$ and $y^2$, and therefore asymptotic expansions of $G_m (1-y^2)$ for large negative $1-y^2$ are useful. Using Eqn.~(C9) and identity {\sl 15.3.7} of Ref.~\cite{AS} we find for large $y^2$
$$
y^2 \,G_m (1-y^2) \rightarrow \frac{(m-1)!\, 2^{m-1}}{(2 m-1)!!\; } - \frac{\pi}{2\, |y|} + {\cal{O}} (1/y^2)\, . \eqno({\rm C15})
$$
Direct expansion of Eqn.~(C10) using Eqn.~(C14) leads to the same result.


\end{document}